\documentclass[5p,times,twocolumn,authoryear]{elsarticle}

\usepackage{ecrc}
\usepackage{amssymb}
\usepackage{amsmath}
\usepackage{booktabs}
\usepackage{float}
\usepackage{balance}
\usepackage{subfig}
\usepackage{algorithm}
\usepackage{algpseudocode}
\usepackage{natbib}
\usepackage[colorlinks,citecolor=blue,urlcolor=blue,linkcolor=blue,bookmarks=false,hypertexnames=true]{hyperref}

\runauth{Flores R. et al.}

\begin{document}
\raggedbottom
\pagestyle{plain}

\begin{frontmatter}

\title{Analysis of Binary O-type Star Systems Using Recurrent Neural Networks}

\author[1]{Miguel Flores R.}
\author[1,2]{Luis J. Corral}
\author[3]{Celia R. Fierro-Santill\'an}
\author[1,2]{Silvana G. Navarro}

\address[1]{Centro Universitario de Ciencias Econ\'omico Administrativas (CUCEA), Universidad de Guadalajara, 45100 Zapopan, Jal., Mexico}
\address[2]{Instituto de Astronom\'ia y Meteorolog\'ia (IAM), Universidad de Guadalajara, 44130 Guadalajara, Jal., Mexico}
\address[3]{Escuela Nacional Colegio de Ciencias y Humanidades, Plantel Sur (ENCCH Sur), Universidad Nacional Aut\'onoma de M\'exico (UNAM), 04500, Mexico City, Mexico}

\begin{abstract}
We examine how adding a secondary contribution to the observed spectrum of an O-type star affects the effective temperature and surface gravity estimated by a recurrent neural network. The network uses equivalent widths measured for 34 optical lines. We modify observed IACOB spectra before normalization in two ways: by adding a blackbody continuum or by adding a second observed spectrum. Across the archived blackbody cases from 15,000 to 25,000~K, the largest within-star spreads in the estimates are 20~K and 0.001~dex. Relative to the original spectra, the largest absolute differences are 4,235~K and 0.333~dex. In the second method, we multiply the secondary spectrum by one of six factors corresponding to 5, 10, 20, 30, 50, and 75 per cent of its original flux before adding it to the primary spectrum. The population analysis covers 192 ordered pairs and 1,152 simulated binary spectra. Effective temperature and surface gravity estimates generally decrease as the secondary contribution increases. The individual comparisons involving HD14947, HD190429, HD48099, and HD206267 show exceptions: some effective temperature estimates fall below those of both isolated spectra, and some surface gravity estimates increase. Comparisons with previously published values depend on the physical parameter and reference used. The network is sensitive to the addition of a secondary spectrum, but these comparisons do not establish a unique secondary contribution for an observed binary.
\end{abstract}

\begin{keyword}
Methods: data analysis \sep Machine learning \sep Stars: binaries \sep Stars: fundamental parameters \sep Astronomical databases: miscellaneous
\end{keyword}

\end{frontmatter}

\section{Introduction}
\label{sec:introduction}

Automated stellar-spectrum analysis is increasingly important as modern surveys expand the number and diversity of available observations. O-type stars contribute to chemical enrichment and radiative and mechanical feedback in galaxies. Accurate estimates of their effective temperatures and surface gravities are important for studies of stellar evolution and population synthesis. Artificial neural networks provide a practical route to estimating physical parameters from these data, particularly when the input representation reduces the computational cost without discarding diagnostically useful line information. In previous work, we developed a recurrent neural network (RNN) that estimates the effective temperature, $T_{\rm eff}$, and surface gravity, $\log g$, of O-type stars from a sequence of 34 equivalent-width (EW) measurements \citep{Flores2023}. Using diagnostic-line measurements reduces the model's dependence on the sampling and resolution of the full spectrum, provided that the diagnostic lines are covered and their EWs are measured consistently.

The network was trained with single-star atmosphere models. This restriction matters because many massive stars belong to binary or multiple systems. When an observed spectrum contains contributions from two or more stars, the absorption and emission lines can differ from those of the isolated primary component. An additional continuum can also change the line depths and equivalent widths. Changes in these EWs can affect the inferred stellar parameters. For example, HD30614 and HD188001, which are included in the observed sample, have been discussed as runaway binary systems \citep{McSwain_2007,Underhill_1996}.

How a single-star model responds to an increasing secondary contribution remains insufficiently explored. The estimates may stay close to those of the primary or move to values that describe neither component. This matters for spectroscopic surveys that may include unresolved binaries.

Using observed spectra from the IACOB database, we compare the effective temperature and surface gravity estimated for an isolated spectrum with the estimates obtained after adding a secondary contribution. The two methods are complementary. The first adds a blackbody continuum, which introduces no spectral lines from the secondary component. The second combines two observed spectra to simulate an unresolved binary and compares the estimates at different secondary contributions. We analyze 192 ordered pairs at six contributions and then examine individual combinations involving HD14947, HD190429, HD48099, and HD206267.

The aim is to examine the behavior and scope of the proposed method when it processes a spectrum modified by a secondary contribution, and to identify the limits of applying a single-star model to observed binaries. This analysis does not provide orbital solutions or separate the observed spectra into their individual components. Both would require observations covering a substantial fraction of the orbit, with adequate radial-velocity sampling. Section~\ref{sec:data} describes the observed spectra and the input to the network, Section~\ref{sec:methods} presents the two methods, Sections~\ref{sec:results} and \ref{sec:cases} show the results and individual comparisons, and Sections~\ref{sec:discussion} and \ref{sec:conclusions} discuss the conclusions and the additional factors required for future work. The appendices provide representative spectra for the blackbody-continuum experiment (\ref{app:blackbody_spectra}), diagnostic-line comparisons for HD48099 and HD206267 (\ref{app:line_profiles}), and a full optical comparison for HD48099 (\ref{app:full_spectra}).

\section{Observed spectra and parameter estimation}
\label{sec:data}

\begin{figure}[!b]
\centering
\subfloat[Stored and normalized IACOB flux for HD14947.\label{fig:iacob_fits}]{\includegraphics[width=0.9\linewidth]{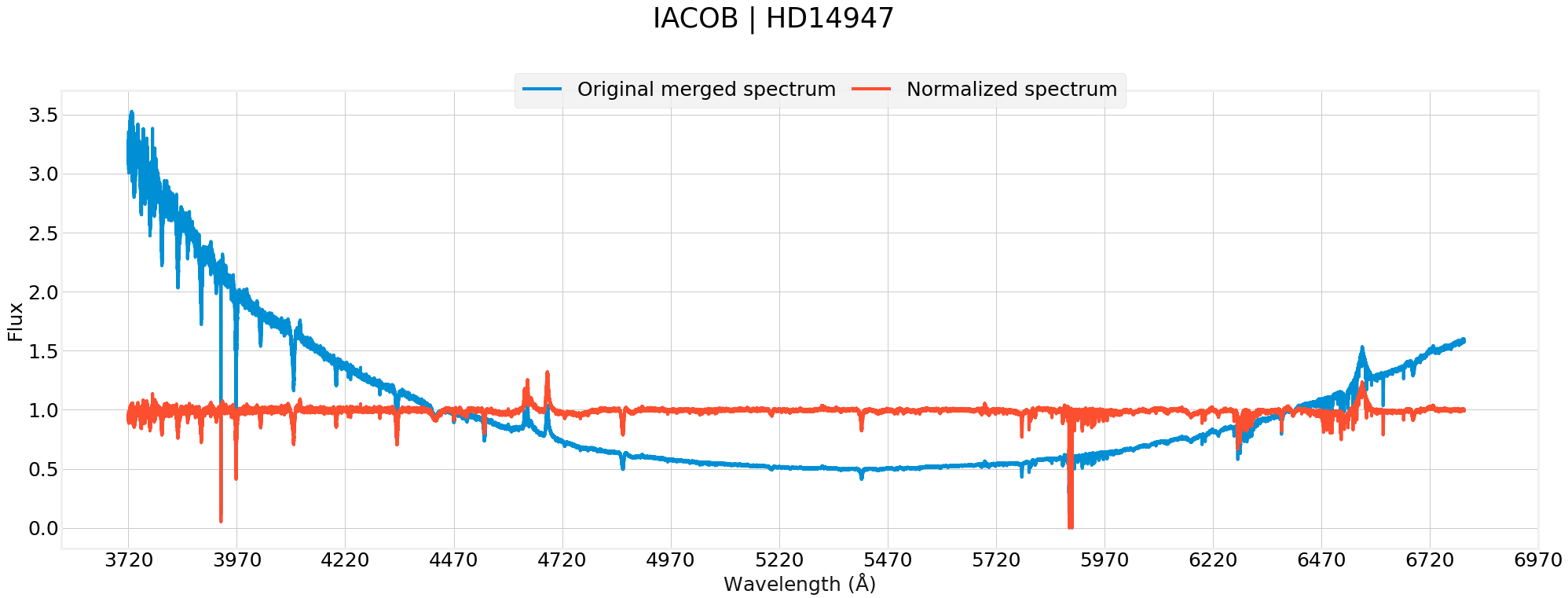}}
\\
\subfloat[Blackbody continuum at 25,000~K.\label{fig:blackbody_curve}]{\includegraphics[width=0.75\linewidth]{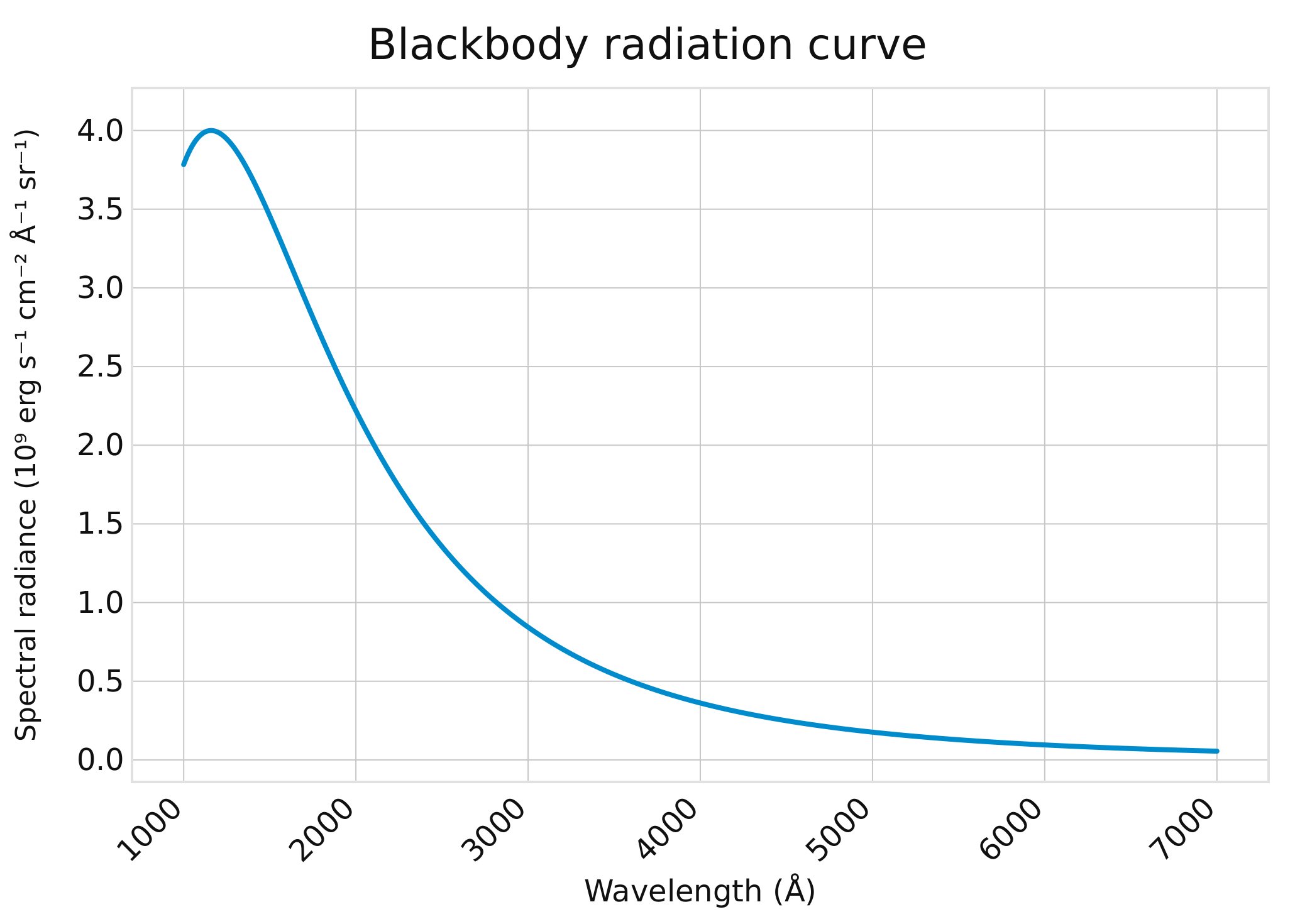}}
\caption{Input representations used to define the continuum experiment. The analysis is restricted to 3700--7000~\AA.}
\label{fig:iacob_blackbody}
\end{figure}

The observed spectra were selected from the IACOB database used to validate the original RNN system \citep{Holgado_2018,Flores2023}. The FITS products provide both an unnormalized stored flux and a normalized spectrum. Figure~\ref{fig:iacob_blackbody}(a) illustrates these two representations for HD14947. The experiments use the optical interval from 3700 to 7000~\AA, matching the region in which the diagnostic EWs are measured.

The initial proposal was to modify the synthetic spectra used to train the network. Those spectra were available only in normalized form, so the addition was instead performed with the observed IACOB spectra, for which the original flux is also available. Combining the spectra before normalization allows the continuum used in the equivalent-width measurements to include both contributions. An equivalent width measures the integrated contrast of a spectral line relative to its local continuum. A change in line depth, width, or profile can therefore change the input to the network, even when the same diagnostic lines are measured.

The RNN accepts an ordered vector of 34 EWs and simultaneously estimates effective temperature, $T_{\rm eff}$, and surface gravity, expressed as $\log g$. Throughout this work, $\log g\equiv\log_{10}[g/(\mathrm{cm\,s^{-2}})]$, where $g$ is the surface acceleration; differences in $\log g$ are reported in dex. The synthetic spectra used to train the network cover approximately $20{,}000\leq T_{\rm eff}\leq58{,}000$~K and $2.4\leq\log g\leq4.2$ \citep{Flores2023}. The simulated binary spectra are obtained by adding the unnormalized observed fluxes and normalizing the resulting spectrum. The 34 EWs are then measured and processed with the same network used for the isolated spectra. This order allows the normalization to include the contributions of both stars.

\section{Methodology}
\label{sec:methods}

\subsection{Addition of a blackbody continuum}
\label{sec:blackbody_method}

The first experiment approximates a companion as a smooth blackbody continuum added to the unnormalized primary spectrum. This construction deliberately excludes the secondary's line spectrum and therefore tests whether continuum dilution alone alters the 34 measured EWs enough to change the RNN output. For a primary flux $F_{\lambda,1}$ and a blackbody continuum $B_\lambda(T_{\rm BB})$, the combined signal can be written schematically as

\begin{equation}
F_{\lambda,\mathrm{BB}} = F_{\lambda,1} + A\,B_\lambda(T_{\rm BB}),
\label{eq:blackbody_sum}
\end{equation}

where $A$ scales the Planck radiance to the units of the stored spectrum. The archival records do not specify $A$ or how it was set across temperatures. The equivalent-width comparison therefore describes the archived spectra and does not establish temperature dependence at a fixed continuum contribution. Algorithm~\ref{alg:blackbody} summarizes the procedure.

\begin{algorithm}[H]
\caption{Blackbody-continuum experiment}
\label{alg:blackbody}
\begin{algorithmic}[1]
\Require Unnormalized observed primary spectrum
\Require Blackbody continuum defined over the observed wavelength range
\State Add the blackbody continuum to the observed flux
\State Normalize the combined spectrum
\State Measure the 34 equivalent widths
\State Process the EW sequence with the RNN
\State \Return Estimated $T_{\rm eff}$ and $\log g$
\end{algorithmic}
\end{algorithm}

Representative spectra modified with a 20,000~K continuum are provided in \ref{app:blackbody_spectra} (Figure~\ref{fig:blackbody_spectra}). Although differences are visible in some wavelength regions, the normalized line profiles remain close to those of the original spectra.

The added blackbody continuum contains no absorption or emission lines. The second method introduces the line spectrum of the secondary star as well as its continuum. The distinction is important for a system whose input consists of equivalent widths: a second observed spectrum can change the morphology of the measured lines, in addition to changing their contrast after normalization.

\subsection{Addition of two observed spectra}
\label{sec:observed_method}

The second method consists of adding two unnormalized observed O-star spectra. The general analysis considers a primary star with a catalog effective temperature greater than or equal to that of the secondary star. The individual comparison in Section~\ref{sec:hd14947} also considers HD30614 as the primary and the hotter HD14947 as the secondary. For a contribution $\alpha$ of the secondary spectrum, the simulated binary spectrum is defined as

\begin{equation}
\begin{split}
F_{\lambda,\mathrm{bin}}(\alpha) &= F_{\lambda,1} + \alpha F_{\lambda,2},\\
\alpha &\in \{0.05,0.10,0.20,0.30,0.50,0.75\}.
\end{split}
\label{eq:binary_sum}
\end{equation}

Here $F_{\lambda,1}$ and $F_{\lambda,2}$ are the original fluxes of the primary and secondary spectra, respectively. The primary spectrum is included at 100 per cent of its original flux, while the secondary spectrum is multiplied by $\alpha$. We refer to $100\alpha$ as the percentage contribution of the secondary spectrum. This definition does not assign the secondary a fraction $\alpha$ of the total combined flux; that fraction is $\alpha F_{\lambda,2}/(F_{\lambda,1}+\alpha F_{\lambda,2})$ and depends on wavelength. The percentages used in this experiment therefore do not determine a physical luminosity ratio for the components. Algorithm~\ref{alg:observed} summarizes the procedure.

In the main experiment, no additional Doppler shift was applied. The line-profile comparison in \ref{app:velocity} shifts the secondary spectrum before adding its flux to the primary. To first order, $\lambda'=\lambda(1+\Delta v/c)$. An imposed shift of $30~\mathrm{km\,s^{-1}}$ moves the secondary H$\alpha$ line at 6563~\AA\ by about $0.66$~\AA. A negative shift moves the line toward shorter wavelengths.

\begin{algorithm}[H]
\caption{Observed-spectrum addition experiment}
\label{alg:observed}
\begin{algorithmic}[1]
\Require Unnormalized observed spectrum of the primary star
\Require Unnormalized observed spectrum of the secondary star
\For{$\alpha$ in $\{0.05,0.10,0.20,0.30,0.50,0.75\}$}
  \State Scale the secondary flux by $\alpha$
  \State Add the scaled secondary to the primary
  \State Normalize the composite spectrum
  \State Measure the 34 equivalent widths
  \State Estimate $T_{\rm eff}$ and $\log g$ with the RNN
\EndFor
\end{algorithmic}
\end{algorithm}

For either parameter $\theta\in\{T_{\rm eff},\log g\}$, we calculate the difference between the RNN estimates for the simulated binary spectrum and the isolated primary spectrum as

\begin{equation}
\Delta\theta(\alpha) = \widehat{\theta}_{1+2}(\alpha)-\widehat{\theta}_{1},
\label{eq:parameter_difference}
\end{equation}

where $\widehat{\theta}_{1+2}$ is the RNN estimate for the composite spectrum and $\widehat{\theta}_{1}$ is the corresponding archived estimate for the isolated primary. Reproducing an isolated estimate requires the same input spectrum, normalization, EW measurements and feature scaling, trained network, and inference settings.

The general comparison in Section~\ref{sec:population_results} corresponds to 192 combinations of primary and secondary spectra. Each combination was evaluated at the six percentages in Equation~\ref{eq:binary_sum}, giving 1,152 simulated binary spectra. The individual comparisons in Section~\ref{sec:cases} use the available estimates for HD190429, HD48099, and HD206267 and the specified combinations with HD14947.\footnote{The archive used for the HD14947 combinations contains 380 ordered pairs and 2,280 simulated spectra, including both orders of addition. The notebook selects 192 combinations from this archive by retaining pairs whose primary is at least as hot as the secondary on the catalog temperature scale.} The continuum, population, and individual-system comparisons use their own archived isolated-star baselines, even where the same star appears in more than one analysis.

\section{Results}
\label{sec:results}

\subsection{Results from the addition of a blackbody continuum}

\begin{figure}[H]
\centering
\includegraphics[width=0.96\linewidth]{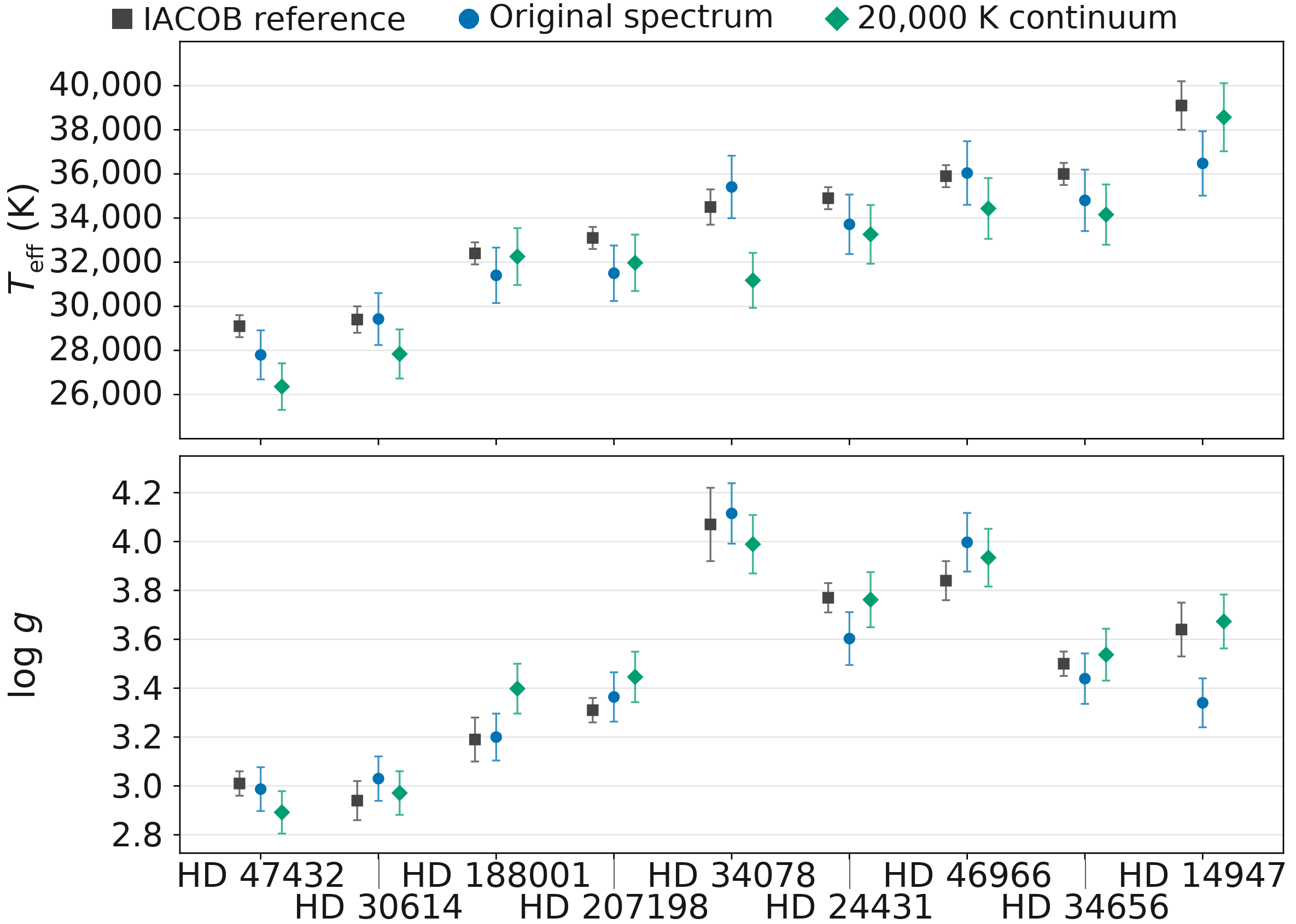}
\caption{RNN estimates of effective temperature (top) and surface gravity (bottom) for the original observed spectra and after adding a 20,000~K blackbody continuum, compared with IACOB reference values. The RNN error bars are $\pm$4\% of the estimated effective temperature and $\pm$3\% of the estimated $\log g$, following the notebook convention. The IACOB bars show the catalog uncertainties. Markers are offset horizontally, and star identifiers are staggered.}
\label{fig:blackbody_results}
\end{figure}

Figure~\ref{fig:blackbody_results} compares the RNN outputs obtained from the original spectra and from spectra modified with a blackbody continuum. Relative to the original spectra, the largest absolute differences among the tested continua are 4,235~K for HD34078 and 0.333~dex for HD14947.

For O-star temperatures, the blackbody maximum lies at wavelengths shorter than the analyzed visible interval, as illustrated in Figure~\ref{fig:blackbody_curve}. Over 3700--7000~\AA\ the added continuum varies smoothly but can still dilute the lines after normalization. This motivates comparison with an observed secondary spectrum, which contributes absorption and emission structure as well as continuum flux.

% Queue the Section 4.2 figures before its text to keep the top floats nearby.
\begin{figure}[!t]
\centering
\subfloat[Estimated Effective Temperature.\label{fig:binary_teff}]{\includegraphics[width=0.85\linewidth]{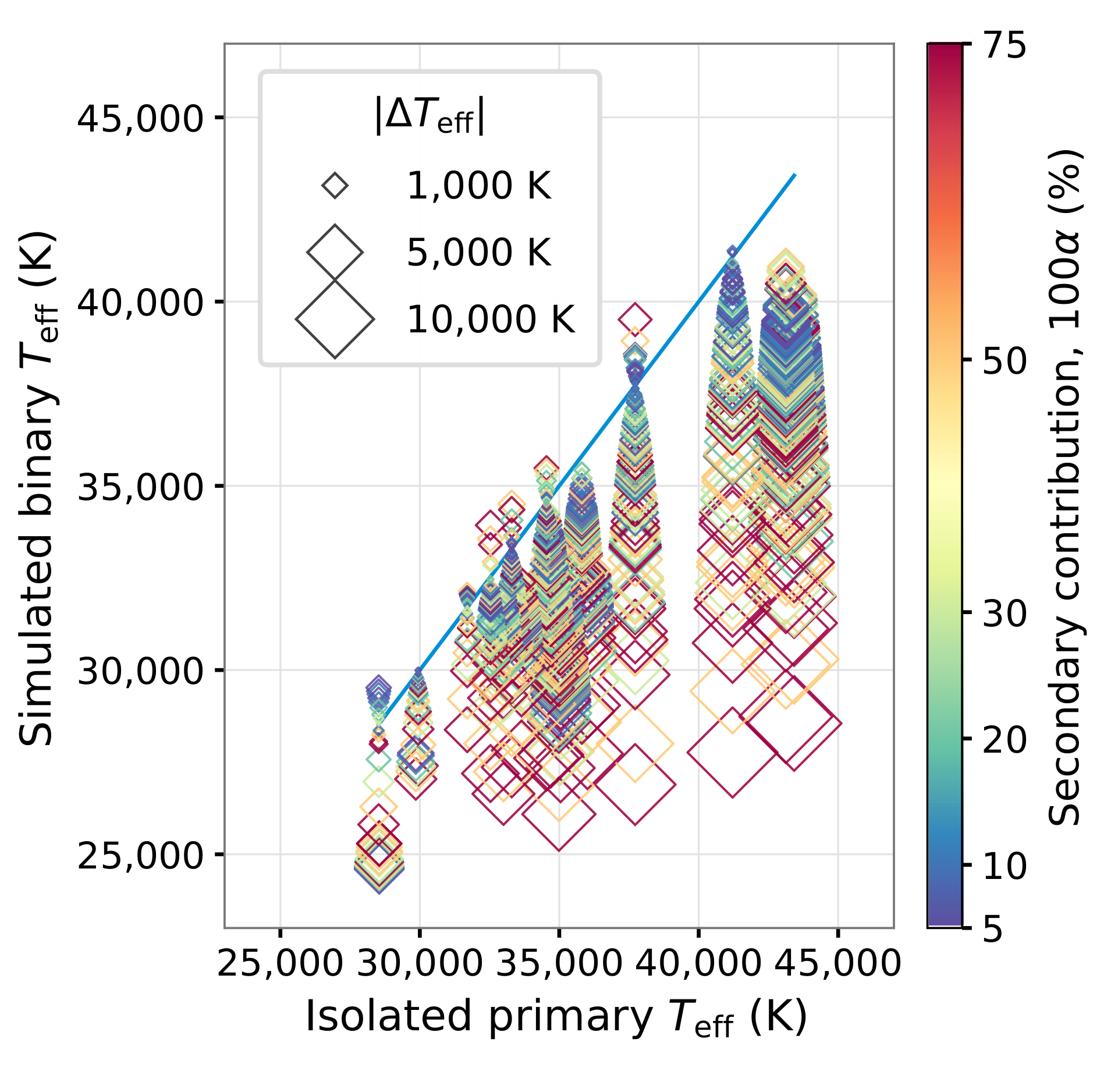}}
\\
\subfloat[Estimated Surface Gravity.\label{fig:binary_logg}]{\includegraphics[width=0.85\linewidth]{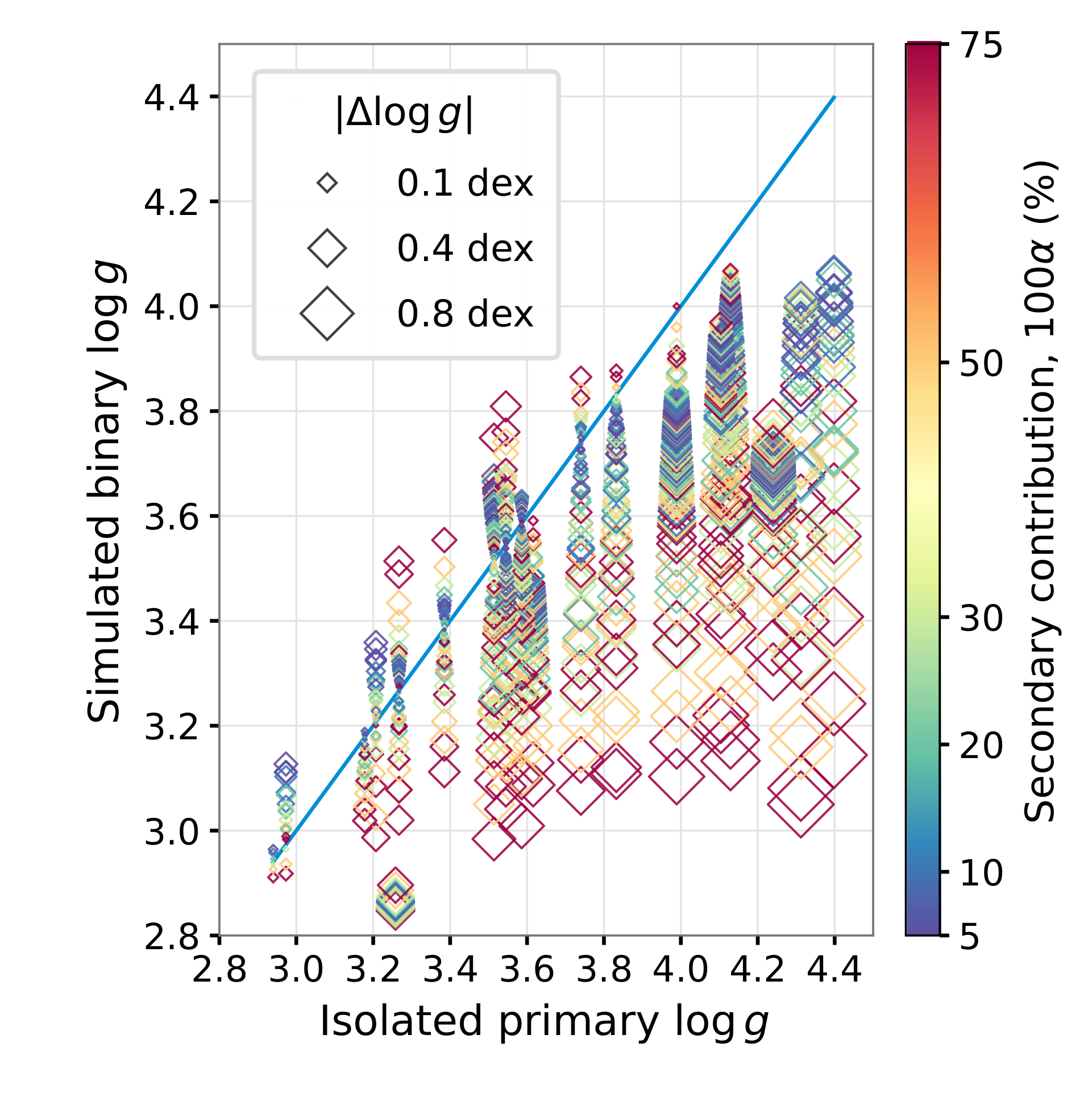}}
\caption{RNN estimates of effective temperature (a) and surface gravity (b) for simulated binary spectra and their isolated primary spectra. Each diamond represents one of 192 ordered pairs at one of six contributions. Color indicates the secondary contribution, $100\alpha$; diamond area represents the magnitude of the difference from the isolated-primary estimate. The diagonal marks equality. Identical coordinates still coincide. The accompanying data identify both components and their catalog parameters.}
\label{fig:binary_estimates}
\end{figure}

\begin{figure*}[!t]
\centering
\includegraphics[width=\textwidth]{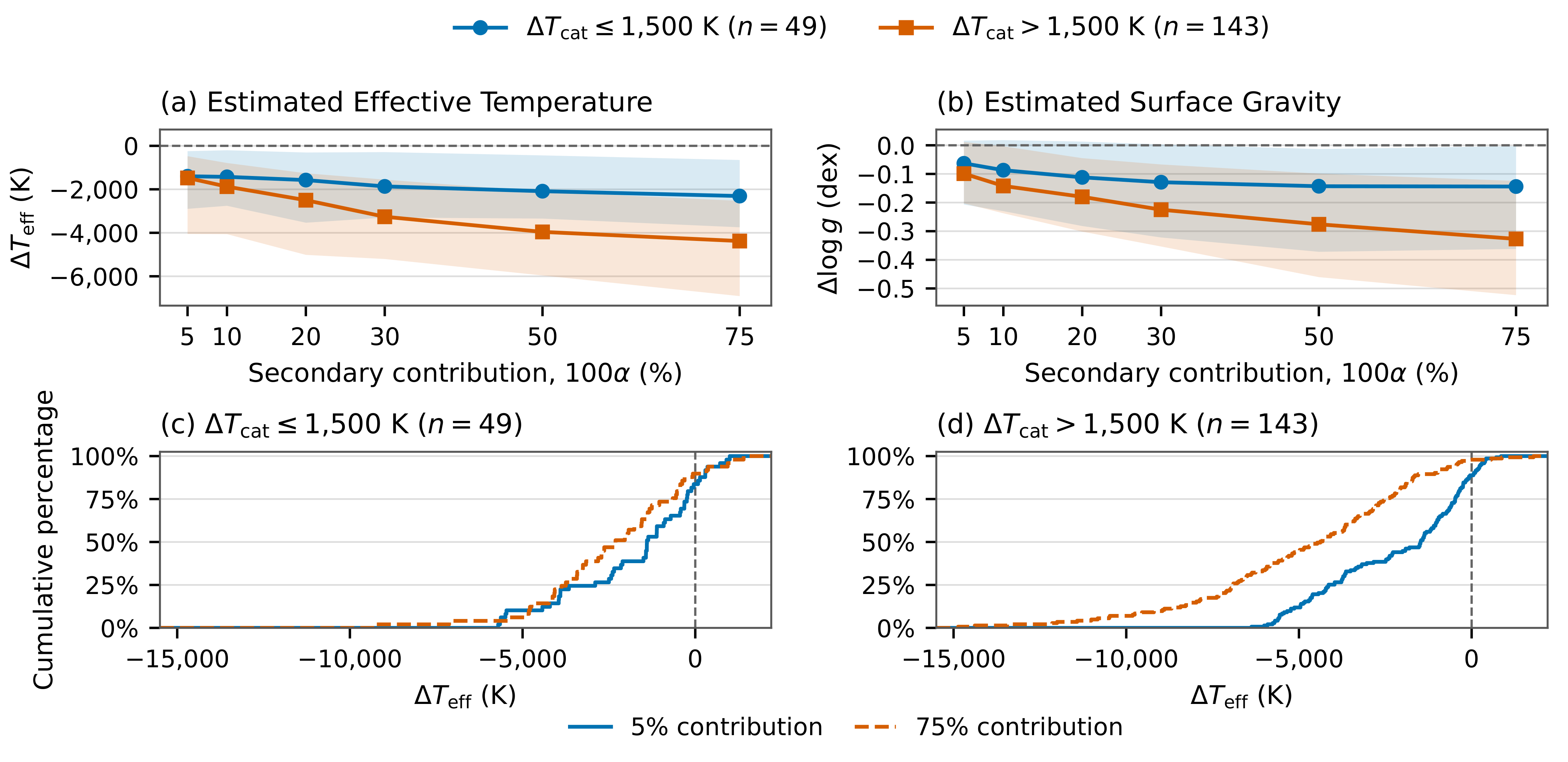}
\caption{Differences between RNN estimates for simulated binary spectra and isolated primary spectra, grouped by catalog effective-temperature contrast $\Delta T_{\rm cat}$. Panels (a) and (b) show the median differences in estimated effective temperature and surface gravity at each secondary contribution; shading spans the 25th--75th percentiles across pairs. Panels (c) and (d) show the empirical cumulative distributions of $\Delta T_{\rm eff}$ at 5 and 75 per cent. At a given $\Delta T_{\rm eff}$, the ordinate is the percentage of pairs with an equal or smaller difference. Dashed gray lines mark zero difference.}
\label{fig:binary_differences}
\end{figure*}

\begin{figure*}[!t]
\centering
\includegraphics[width=\textwidth]{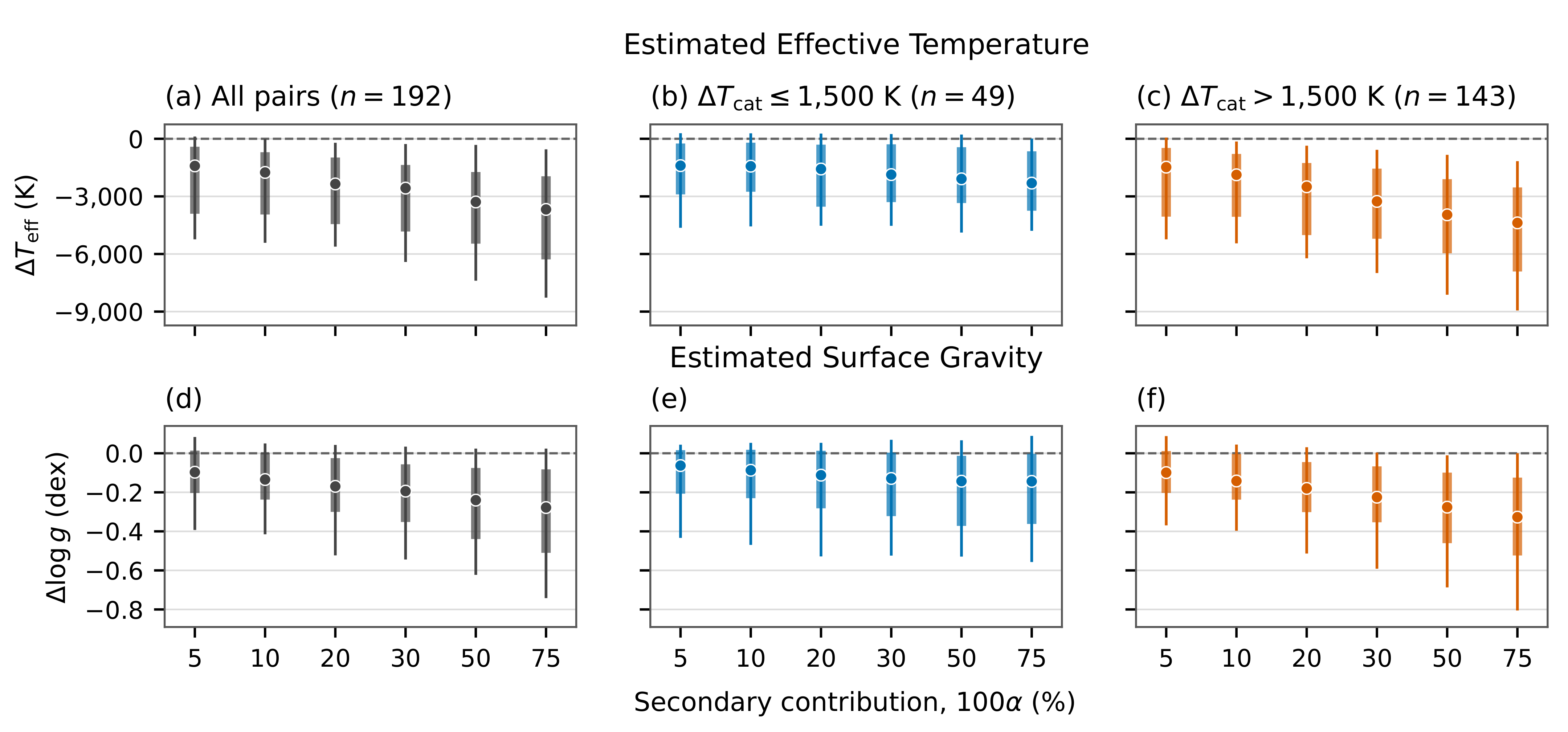}
\caption{Distributions of the differences in estimated effective temperature (a--c) and surface gravity (d--f) at the six tested secondary contributions, shown at equal horizontal intervals. The columns show all 192 pairs and the groups with catalog effective-temperature contrasts at or below and above 1,500~K. Dots mark medians, thick intervals span the 25th--75th percentiles, and thin intervals span the 10th--90th percentiles across pairs. The vertical scale is common within each row. Dashed lines mark zero difference.}
\label{fig:population_distributions}
\end{figure*}

Across the nine stars, the archived cases at 15,000, 20,000, and 25,000~K give a maximum within-star spread of 20~K in estimated effective temperature and 0.001~dex in estimated $\log g$. These differences are much smaller than the RNN error bars in Figure~\ref{fig:blackbody_results}, so only the 20,000~K case is plotted. This does not imply that adding a continuum has no effect.

The small spread applies to these archived continuum cases and does not establish that an added continuum leaves equivalent widths unchanged in general. The amount of dilution depends on the continuum contribution relative to the primary spectrum.

\subsection{Results from the addition of two observed spectra}
\label{sec:population_results}

Figure~\ref{fig:binary_estimates} compares the effective temperature and surface gravity estimates for the simulated binary spectra with those for the isolated primary spectra. Most estimates lie below the line of equality, and the differences generally increase in magnitude with the contribution of the secondary component. The estimates for a simulated binary spectrum need not lie between the separate estimates for the two component spectra (Section~\ref{sec:cases}).

We express the differences between the RNN estimates as $\Delta T_{\rm eff}$ in kelvin and $\Delta\log g$ in dex (Equation~\ref{eq:parameter_difference}). Negative differences indicate lower estimates for the simulated binary spectrum; zero indicates agreement with the isolated-primary estimate. The catalog effective-temperature contrast between the components, $\Delta T_{\rm cat}=T_{\rm eff,1}^{\rm cat}-T_{\rm eff,2}^{\rm cat}$, is nonnegative by construction. This selection imposes no ordering in surface gravity. Equal catalog effective-temperature contrasts also need not give equal RNN estimates: at 75 per cent, HD24431 with HD36486 or HD188209 has the same contrast of 4,800~K but gives $\Delta T_{\rm eff}=-5{,}934$~K or $-1{,}932$~K, respectively.

Figures~\ref{fig:binary_differences} and \ref{fig:population_distributions} compare 49 pairs with $\Delta T_{\rm cat}\leq1{,}500$~K and 143 pairs with $\Delta T_{\rm cat}>1{,}500$~K. At 75 per cent, the median differences in estimated effective temperature are $-2{,}309$~K and $-4{,}379$~K in the two groups, respectively; the median differences in estimated surface gravity are $-0.144$ and $-0.327$~dex. The cumulative distributions for effective temperature overlap at 5 and 75 per cent in both groups, especially in the smaller-contrast group. Thus a 5 per cent contribution in one pair can give a difference comparable to that for a larger contribution in another pair.

Across all 192 pairs, the median $\Delta T_{\rm eff}$ is approximately $-1{,}413$~K at 5 per cent and $-3{,}680$~K at 75 per cent; the median $\Delta\log g$ is $-0.097$ and $-0.278$~dex, respectively. Figure~\ref{fig:population_distributions} shows the spread around these medians. The percentile intervals for the two groups overlap even at 75 per cent. They describe variation among the simulated pairs, not RNN estimation uncertainties or confidence intervals. The same ordered pairs recur at all six contributions.

At 75 per cent, the group at or below the catalog cutoff has smaller median decreases in both estimates for each of the four tested cutoffs. The cutoffs of 1,000, 1,500, 2,000, and 3,000~K place 38, 49, 61, and 90 pairs at or below the cutoff, respectively. The adopted 1,500~K division is comparable to the scale of the RNN's effective-temperature uncertainty \citep{Flores2023}; it is a descriptive division of the sample, not a physical boundary or a cutoff chosen to optimize the separation.

% The two-column case-study text may fill space below the wide population figures.
\section{Analysis of individual binary systems}
\label{sec:cases}

% Queue the five case-study panels as wide top floats, in the text's order.
\begin{figure*}[!t]
\centering
\includegraphics[width=0.87\textwidth]{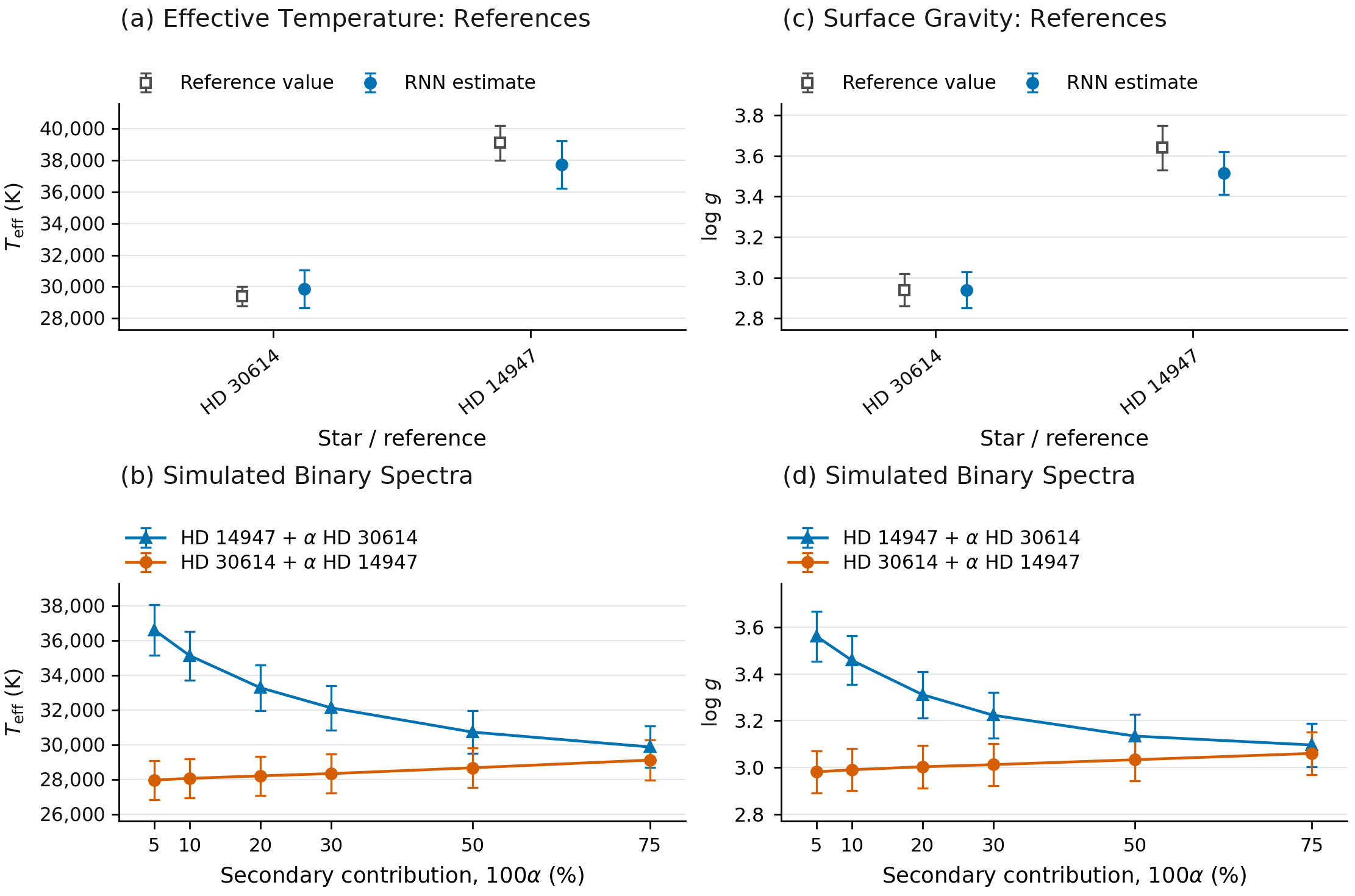}
\caption{Estimated effective temperature and surface gravity for combinations of HD14947 and HD30614, with each star considered in turn as the primary. Panels (a,c) compare the estimates for the isolated observed spectra with the IACOB reference values \citep{Holgado_2018}. Panels (b,d) show the estimates after adding 5, 10, 20, 30, 50, and 75 per cent of the secondary spectrum. The primary spectrum is included at its full original flux in both orders of addition.}
\label{fig:hd14947_pair}
\end{figure*}

\begin{figure*}[!t]
\centering
\includegraphics[width=0.87\textwidth]{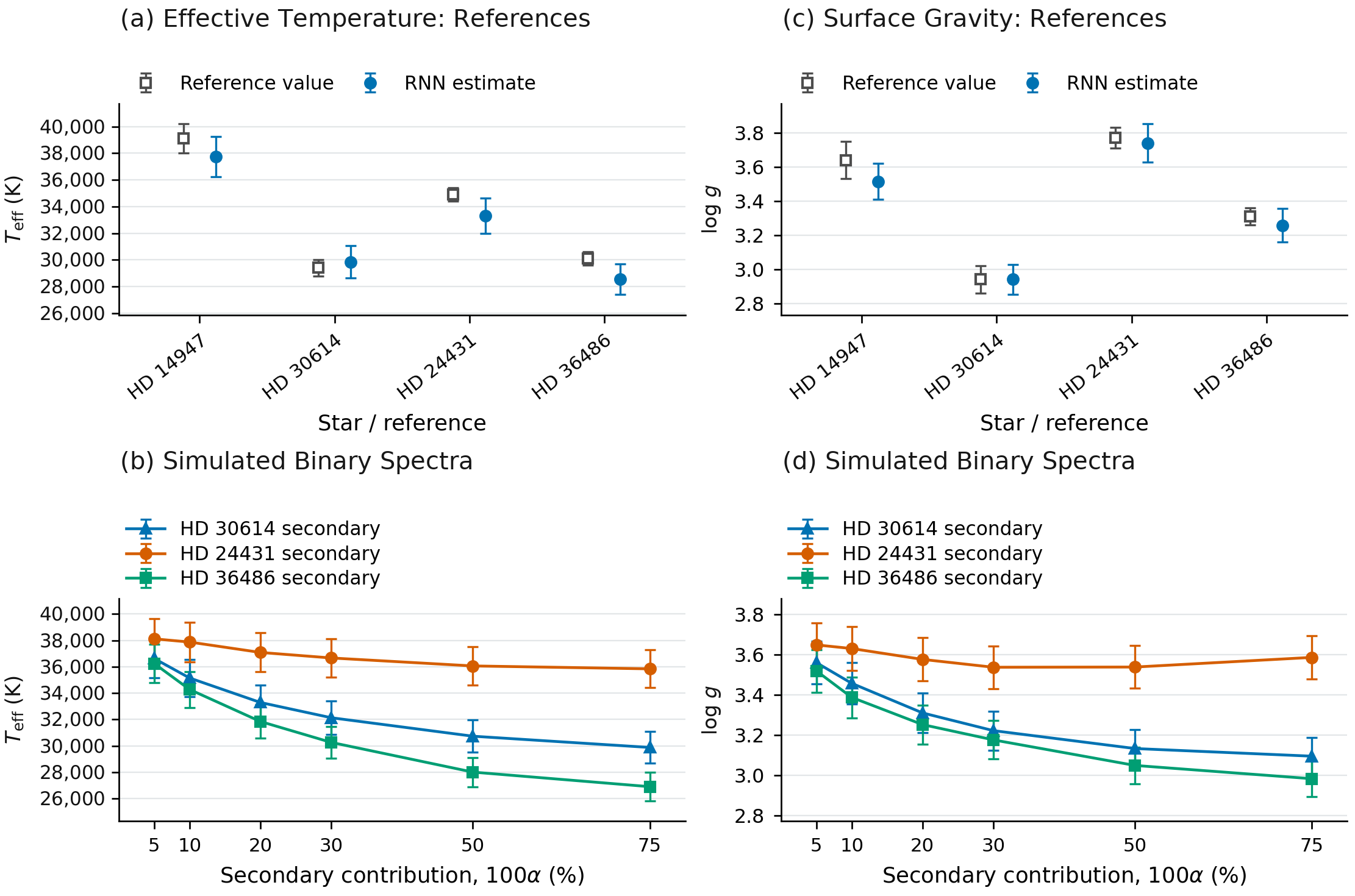}
\caption{Estimated effective temperature and surface gravity for simulated binary spectra using HD14947 as the primary and HD30614, HD24431, or HD36486 as the secondary. Panels (a,c) compare the estimates for the isolated observed spectra with the IACOB reference values \citep{Holgado_2018}. Panels (b,d) show the estimates at the six percentages of secondary contribution. Each curve identifies the star used as the secondary component.}
\label{fig:hd14947_multi}
\end{figure*}

\begin{figure*}[!t]
\centering
\includegraphics[width=0.87\textwidth]{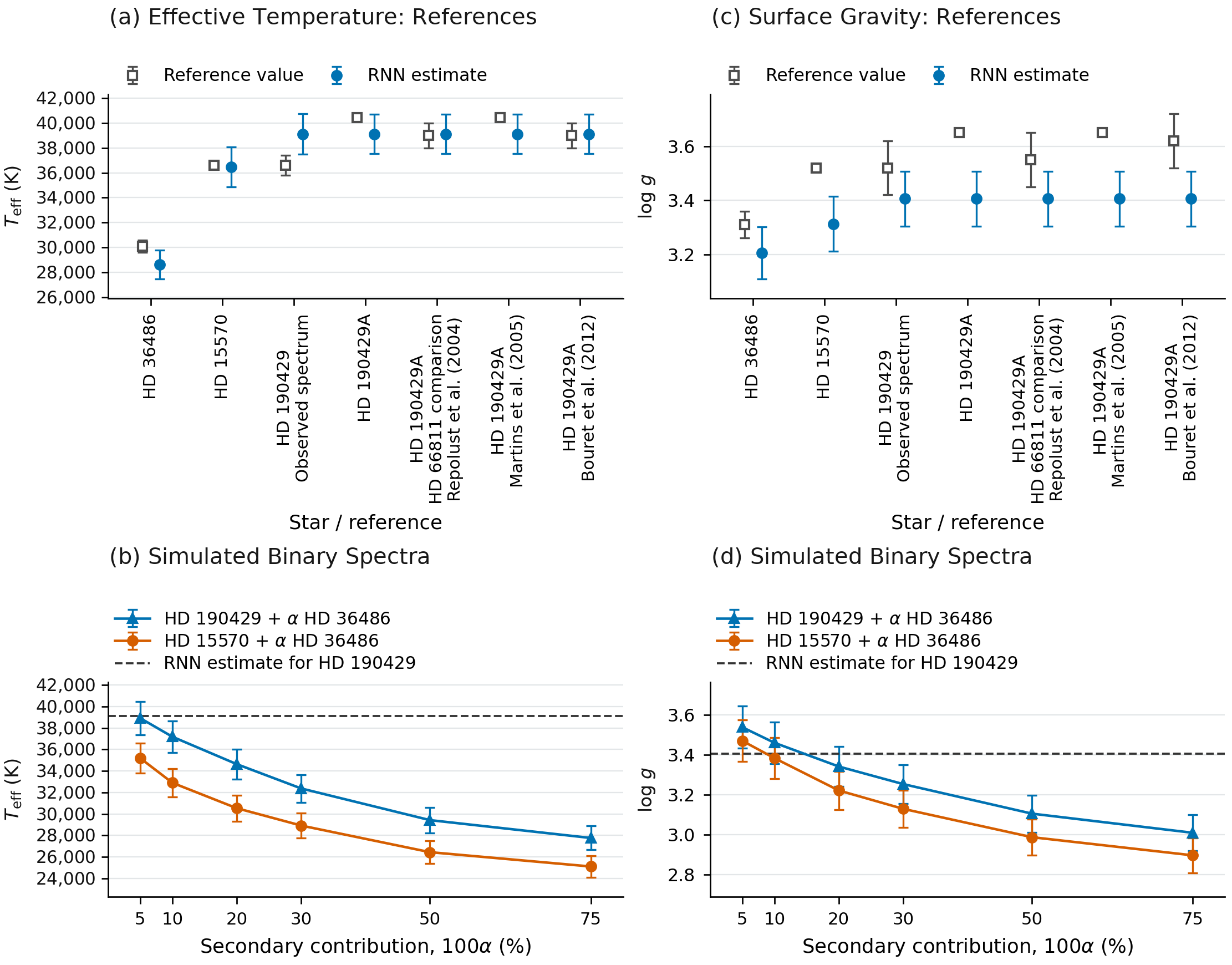}
\caption{Estimated effective temperature and surface gravity for HD190429 and for simulations using HD15570 or the observed HD190429 spectrum as the primary, with HD36486 as the secondary. Panels (a,c) compare the RNN estimates with previously published values \citep{Holgado_2018,2004,2012,Martins}. The estimate from the unresolved HD190429 spectrum is repeated beside the primary-component references; the \citet{2004} reference is for HD66811. Panels (b,d) show RNN estimates for the combined spectra at different percentages of contribution. Dashed lines indicate the estimates obtained from the unresolved HD190429 spectrum before adding HD36486.}
\label{fig:hd190429_results}
\end{figure*}

\begin{figure*}[!t]
\centering
\includegraphics[width=0.87\textwidth]{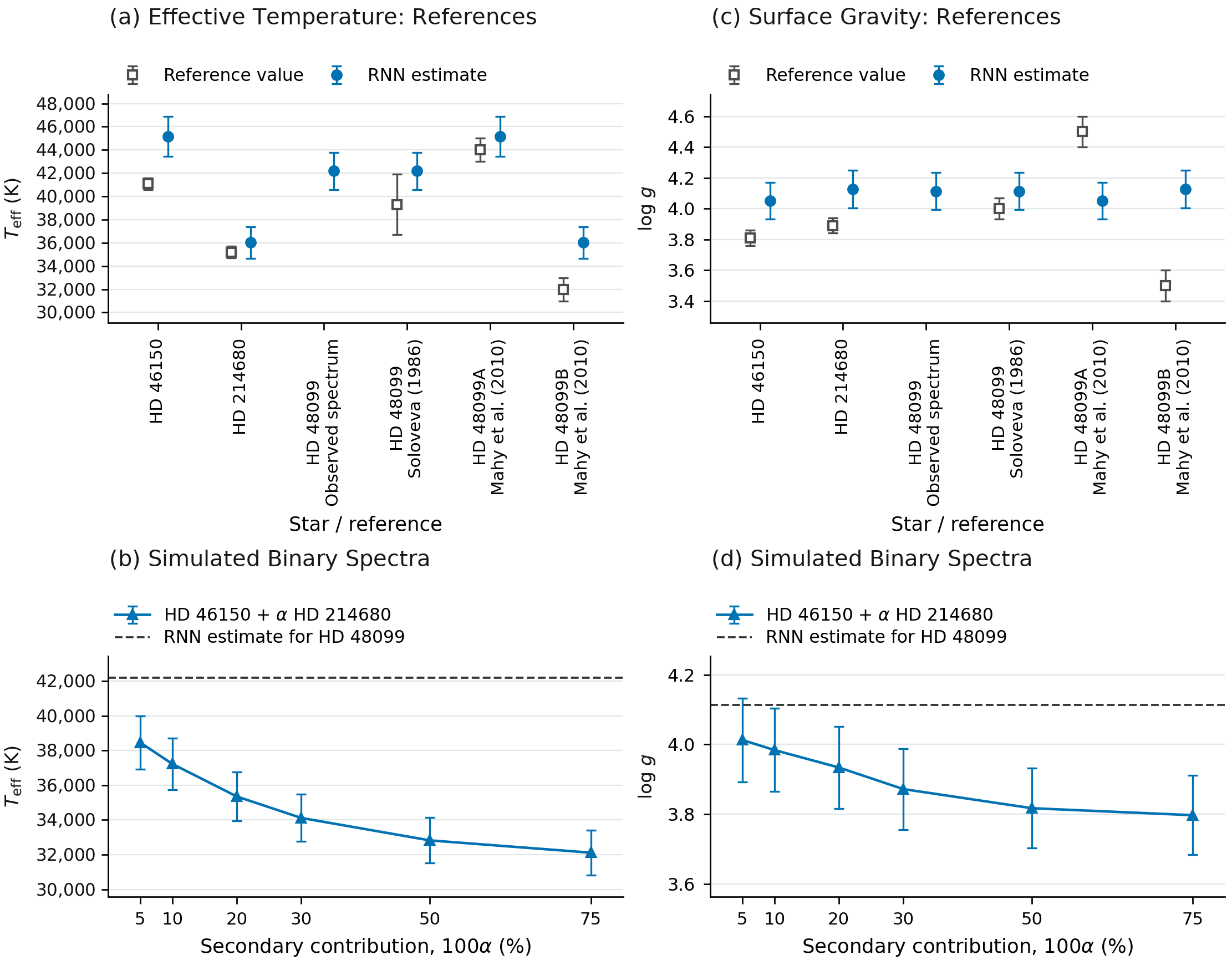}
\caption{Estimated effective temperature and surface gravity for HD48099 and for simulations using HD46150 as the primary component and HD214680 as the secondary component. Panels (a,c) compare estimates from the individual observed spectra with the reference values for HD48099 \citep{Soloveva_1986} and its components \citep{Mahy_2010}. The estimates beside the component references correspond to HD46150 and HD214680. Panels (b,d) show estimates for the simulated binary spectra at different percentages of contribution. Dashed lines indicate the estimates obtained from the observed spectrum of HD48099.}
\label{fig:hd48099_results}
\end{figure*}

\begin{figure*}[!t]
\centering
\includegraphics[width=0.87\textwidth]{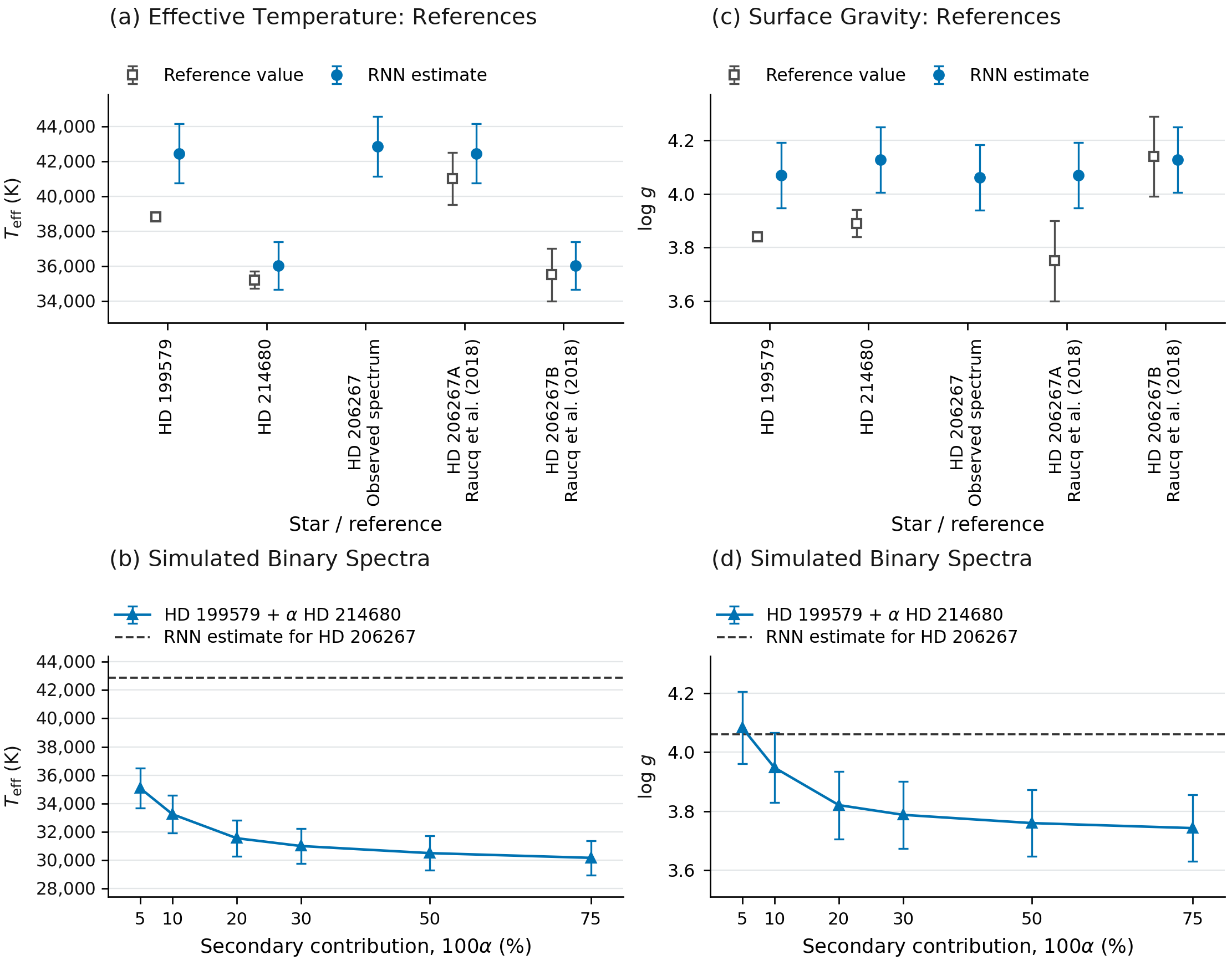}
\caption{Estimated effective temperature and surface gravity for HD206267 and for simulations using HD199579 as the primary component and HD214680 as the secondary component. Panels (a,c) compare the estimates from individual observed spectra with the component reference values from \citet{Raucq2018}. The estimates beside the component references correspond to HD199579 and HD214680. Panels (b,d) show estimates for the simulated binary spectra at different percentages of contribution. Dashed lines indicate the estimates obtained from the observed spectrum of HD206267.}
\label{fig:hd206267_results}
\end{figure*}

In the following examples, we use observed spectra of stars with spectral types and luminosity classes similar to those of the components of the binary systems. Table~\ref{tab:simulation_components} identifies the spectra used for the three observed binary systems. This selection provides examples with comparable spectral classifications; it does not require the selected stars to have the same atmospheric parameters as the actual binary components. 

Figures~\ref{fig:hd14947_pair}--\ref{fig:hd206267_results} compare the estimated effective temperature and surface gravity with previously published values and show how the estimates change with the contribution of the secondary spectrum. Panels (a) and (c) show the reference values and the estimates from individual observed spectra; panels (b) and (d) show the estimates for the simulated binary spectra. Effective temperature is plotted in kelvin (K), and surface gravity is expressed as $\log g$ with the definition given in Section~\ref{sec:data}. The error bars show the uncertainties supplied with the estimates and reference values. Where a reference has no error bar, an uncertainty was not provided.

Unless another source is identified, the reference values are those published by the IACOB project \citep{Holgado_2018}. The HD190429A reference without an author label is the value assigned from the spectral-type and luminosity-class calibration of \citet{Martins}. The estimate from the observed spectrum of HD190429 is repeated for comparison with the published primary-component values. For HD48099 and HD206267, the estimates shown beside the published component values are those of the individual stars used to simulate the components. They should not be interpreted as separate estimates from disentangled spectra of the binary components. The entries for the observed spectra of HD48099 and HD206267 show only the estimates of the proposed system, without an additional independent reference value. In the analysis below, the closest value denotes the smallest absolute difference from a specified reference among the six percentages considered, separately for each physical parameter. These comparisons do not establish a simultaneous fit of effective temperature and surface gravity.

Panels (b) and (d) contain one estimate for each of the six contributions; the lines connect the calculated points to make the comparison easier to follow. They do not represent additional calculations at intermediate percentages. The isolated spectra provide the comparison without an added secondary contribution, while the dashed lines in the HD190429, HD48099, and HD206267 figures identify the estimates obtained from their observed spectra.

\begin{table}[H]
\centering
\caption{Observed spectra used to simulate the three binary systems. In each combination the primary spectrum is included at its original flux and the secondary spectrum is multiplied by $\alpha$. The second HD190429 combination adds a secondary spectrum directly to the observed spectrum of that system.}
\label{tab:simulation_components}
\begin{tabular}{lll}
\toprule
System & Primary & Secondary \\
\midrule
HD190429 & HD15570 & HD36486 \\
HD190429 & HD190429 & HD36486 \\
HD48099 & HD46150 & HD214680 \\
HD206267 & HD199579 & HD214680 \\
\bottomrule
\end{tabular}
\end{table}

\subsection{Simulations using HD14947}
\label{sec:hd14947}

HD190429 is classified as O4If+O9.5II--III, with reference values of $T_{\rm eff}=36{,}600$~K and $\log g=3.52$ published by IACOB \citep{Holgado_2018}. The observed sample includes stars with characteristics similar to those of its components. For an initial comparison, HD14947 represents the hotter primary and HD30614 provides a cooler luminous secondary spectrum. We use HD14947 (O4.5If), with reference effective temperature $T_{\rm eff}=39{,}100$~K and surface gravity $\log g=3.64$, and HD30614 (O9Ia), with $T_{\rm eff}=29{,}400$~K and $\log g=2.94$.

Figure~\ref{fig:hd14947_pair}(b) shows that, when HD14947 is taken as the primary star, the estimated effective temperature decreases from $T_{\rm eff}=36{,}624$~K to $T_{\rm eff}=29{,}872$~K as the contribution of HD30614 increases from 5 to 75 per cent. When HD30614 is taken as the primary and the hotter HD14947 spectrum is added, the estimated effective temperature increases from $T_{\rm eff}=27{,}961$~K to $T_{\rm eff}=29{,}113$~K over the same percentages. Nevertheless, these estimates remain below $T_{\rm eff}=29{,}855$~K, obtained from the isolated HD30614 spectrum. An increase between the 5 and 75 per cent cases therefore does not imply an increase relative to the isolated primary spectrum.

The estimated surface gravity shows similar differences between the two orders of addition (Figure~\ref{fig:hd14947_pair}(d)). With HD14947 as the primary, the estimates decrease from $\log g=3.561$ to $\log g=3.096$ between 5 and 75 per cent. With HD30614 as the primary, they increase from $\log g=2.981$ to $\log g=3.060$. The 5 per cent estimate for HD14947+HD30614 is above the isolated HD14947 estimate of $\log g=3.514$, despite the decreasing trend at larger contributions.

Two additional stars with spectral types and luminosity classes similar to the secondary component of HD190429 are listed in Table~\ref{tab:secondary_analogues}. Figure~\ref{fig:hd14947_multi} shows the results obtained with HD14947 as the primary and HD30614, HD24431, or HD36486 as the secondary. Between the 5 and 75 per cent cases, the estimated effective temperature decreases from $T_{\rm eff}=38{,}117$~K to $T_{\rm eff}=35{,}840$~K when HD24431 is added, and from $T_{\rm eff}=36{,}244$~K to $T_{\rm eff}=26{,}894$~K when HD36486 is added (panel b). In the latter case, the final estimate is below $T_{\rm eff}=28{,}545$~K, obtained for the isolated HD36486 spectrum.

\begin{table}[H]
\centering
\caption{Additional stars used as the secondary component in combinations with HD14947. Spectral type, luminosity class, effective temperature, and surface gravity are the IACOB reference values \citep{Holgado_2018}. Surface gravity is expressed as $\log g$, using $g$ in $\mathrm{cm\,s^{-2}}$.}
\label{tab:secondary_analogues}
\begin{tabular}{lccc}
\toprule
Star & SpT/LC & $T_{\rm eff}$ (K) & $\log g$ \\
\midrule
HD24431 & O9III & 34,900 & 3.77 \\
HD36486 & O9.5IINwk & 30,100 & 3.31 \\
\bottomrule
\end{tabular}
\end{table}

For surface gravity, the estimates obtained by adding HD24431 remain above the isolated HD14947 estimate of $\log g=3.514$. The estimated surface gravity reaches $\log g=3.538$ at a contribution of 30 per cent and then increases to $\log g=3.587$ at 75 per cent (Figure~\ref{fig:hd14947_multi}(d)). These results show that the behavior depends on the secondary spectrum as well as its percentage contribution. A decrease in estimated effective temperature is not necessarily accompanied by a decrease in estimated surface gravity.

\subsection{HD190429}
\label{sec:hd190429}

To complement the previous analysis, we simulate the binary system HD190429 using HD15570 as the primary component and HD36486 as the secondary component. We also apply the same procedure to the observed IACOB spectrum of HD190429, adding HD36486 at the six percentages of contribution. Figure~\ref{fig:hd190429_results}(a,c) compares the estimated effective temperature and surface gravity with the IACOB values \citep{Holgado_2018}, the calibration of \citet{Martins}, and the values for HD190429A published by \citet{2012}. The reference from \citet{2004} corresponds to HD66811, which is included as a comparison star rather than as a direct determination for HD190429A.

For the observed spectrum of HD190429, the proposed system estimates an effective temperature of $T_{\rm eff}=39{,}124$~K and a surface gravity of $\log g=3.406$. The corresponding IACOB reference values are $T_{\rm eff}=36{,}600$~K and $\log g=3.52$. The estimated effective temperature is closer to the primary-component value of $T_{\rm eff}=39{,}000$~K published by \citet{2012}. When HD36486 is added to the HD190429 spectrum, a contribution of 10 per cent gives $T_{\rm eff}=37{,}186$~K, the closest estimate to the IACOB effective temperature among the percentages considered, with a difference of 586~K (panel b).

For surface gravity, the estimate from the HD190429 spectrum is below the reference values shown in panel (c). Adding HD36486 gives $\log g=3.538$ at 5 per cent and $\log g=3.460$ at 10 per cent (panel d). The 5 per cent estimate is closest to the IACOB value of $\log g=3.52$, differing by 0.018~dex. The alternative simulation using HD15570+HD36486 gives $T_{\rm eff}=35{,}194$~K and $\log g=3.470$ at 5 per cent; the estimates of both parameters decrease at larger contributions. Thus the closest percentage depends on the parameter and the spectra used. These comparisons suggest that the observed spectrum may be dominated by the primary component. They do not determine the contribution of the secondary component already present in the observation.

\subsection{HD48099}
\label{sec:hd48099}

The second case is the binary system HD48099 (O5.5V((f))z+O9V), simulated using HD46150 as the primary component and HD214680 as the secondary component (Figure~\ref{fig:hd48099_results}). For the isolated HD46150 spectrum, the estimated effective temperature is $T_{\rm eff}=45{,}183$~K, close to $T_{\rm eff}=44{,}000$~K published by \citet{Mahy_2010} for HD48099A. For HD214680, the estimated effective temperature is $T_{\rm eff}=36{,}020$~K, above the value of $T_{\rm eff}=32{,}000$~K published for HD48099B (panel a). These estimates correspond to the individual spectra used to simulate the components. HD214680 was selected because its O9V classification is the same as that assigned to HD48099B in this comparison. Its IACOB reference effective temperature, $T_{\rm eff}=35{,}200$~K, is nevertheless different from the value published for HD48099B. This difference illustrates why matching spectral type and luminosity class does not make the selected spectrum an exact representation of a binary component. The differences are also apparent for surface gravity: HD46150 gives $\log g=4.052$ and HD214680 gives $\log g=4.127$, compared with $\log g=4.5$ for HD48099A and $\log g=3.5$ for HD48099B from \citet{Mahy_2010} (panel c).

The estimated effective temperature of the simulated binary spectrum decreases from $T_{\rm eff}=38{,}458$~K at 5 per cent to $T_{\rm eff}=32{,}112$~K at 75 per cent (panel b). At 5 per cent, the estimate differs by 842~K from $T_{\rm eff}=39{,}300$~K, the reference attributed to \citet{Soloveva_1986}. Processing the observed HD48099 spectrum gives $T_{\rm eff}=42{,}179$~K, which differs from the estimate for the 5 per cent simulation by 3,721~K. Thus agreement with the published reference is different from agreement with the estimate obtained from the observed spectrum using the proposed system.

For surface gravity, the 5 per cent simulation gives $\log g=4.013$, close to the reference value of $\log g=4.0$ attributed to \citet{Soloveva_1986}. The estimate for the observed HD48099 spectrum is $\log g=4.113$ (panel d). Increasing the contribution of HD214680 decreases both estimated parameters of the simulated binary spectrum, increasing their differences from the estimates for the observed HD48099 spectrum. The agreement with the values of \citet{Soloveva_1986} therefore does not establish a unique contribution of the secondary component. The line profiles in \ref{app:line_profiles} (Figure~\ref{fig:hd48099_lines}) compare the observed HD48099 spectrum with the 5 per cent HD46150+HD214680 simulation and the isolated spectra used to construct it. The broader optical comparison in \ref{app:full_spectra} (Figure~\ref{fig:hd48099_full}) allows differences outside the selected diagnostic lines to be examined.

\subsection{HD206267}
\label{sec:hd206267}

The third case is the binary system HD206267 (O6.5V((f))+O9V), simulated using HD199579 as the primary component and HD214680 as the secondary component (Figure~\ref{fig:hd206267_results}). For the observed HD206267 spectrum, the proposed system estimates an effective temperature of $T_{\rm eff}=42{,}856$~K and a surface gravity of $\log g=4.061$. For the isolated HD199579 spectrum, the estimates are $T_{\rm eff}=42{,}441$~K and $\log g=4.070$. The estimates for these two observed spectra differ by 415~K in effective temperature and 0.009~dex in surface gravity. The primary-component reference in panel (a), from \citet{Raucq2018}, is $T_{\rm eff}=41{,}000$~K. As in the previous example, the estimates beside the published component values are those of the individual stars used in the simulation.

As in the HD48099 example, adding the secondary spectrum produces a rapid decrease in estimated effective temperature. At a contribution of 5 per cent, the simulated HD199579+HD214680 spectrum gives $T_{\rm eff}=35{,}081$~K, which is 7,775~K below the estimate for the observed HD206267 spectrum (panel b). The estimated effective temperature decreases further at larger contributions, so none of the six simulations reproduces the value estimated from the observed HD206267 spectrum. The similarity between the estimates for HD206267 and the isolated HD199579 spectrum is consistent with the primary component dominating the observation. However, contributions below 5 per cent were not evaluated, and the stars used to simulate the system are not its actual components. This comparison therefore does not establish an upper limit on the contribution of the secondary component.

For surface gravity, the percentage that gives the closest estimate depends on the reference considered (panel d). The 5 per cent simulation gives $\log g=4.083$, closest to $\log g=4.061$ estimated for the observed HD206267 spectrum. At 20 per cent, the simulation gives $\log g=3.821$, closest to the IACOB reference for HD199579, $\log g=3.84$. The primary-component reference of $\log g=3.75$ from \citet{Raucq2018} is instead approached at 50 and 75 per cent, with estimates of $\log g=3.760$ and $\log g=3.743$, respectively. These latter simulations have effective temperatures well below the estimate for the observed HD206267 spectrum. Agreement in surface gravity alone therefore does not establish agreement in both physical parameters or support a 20 per cent contribution for HD206267.

The diagnostic lines in \ref{app:line_profiles} (Figure~\ref{fig:hd206267_lines}) show the HD199579+HD214680 simulation at a contribution of 10 per cent. This example illustrates the changes in the spectrum produced by adding the secondary component; it was not selected by simultaneous agreement in effective temperature and surface gravity. We did not use the HD206267 observation (S/N $= 58$) as a full-spectrum reference because the noise, especially at shorter wavelengths, limits a reliable comparison of the line profiles.

\section{Discussion and limitations}
\label{sec:discussion}

\citet{Sana2012} used multi-epoch spectroscopy to estimate that more than 70 per cent of Galactic O-type stars will exchange mass with a companion during their lifetimes. This highlights the importance of accounting for stellar multiplicity in spectroscopic studies of massive stars.

The two methods considered in this work produce different changes in the observed spectrum. Adding a blackbody continuum modifies the unnormalized flux and can dilute the normalized absorption lines. The estimates for the three archived continuum cases differ much less from each other than they do from estimates for the original spectra. Adding an observed secondary spectrum also introduces absorption and emission lines, changing the equivalent widths used by the RNN. The differences between the estimates for isolated and simulated binary spectra show that the network is sensitive to this contribution, even though it was trained with single-star models.

For a simulated binary spectrum, the effective temperature and surface gravity estimated by the network need not represent either component or an average of the isolated-component estimates. The HD30614+HD14947 and HD14947+HD36486 examples give effective temperature estimates below those obtained for both isolated spectra. The surface gravity estimate for HD14947+HD24431 increases at larger contributions after an initial decrease. The trends in Figures~\ref{fig:binary_estimates}--\ref{fig:population_distributions} therefore do not describe every individual pair. The catalog effective-temperature contrast helps characterize the sample but does not determine the estimates for an individual pair.

The comparison with other authors also requires distinguishing the reference for each physical parameter. For HD190429, the percentages that give the closest effective temperature and surface gravity to the IACOB values are different. For HD48099, the 5 per cent simulation gives estimates close to the values attributed to \citet{Soloveva_1986}, while the estimated effective temperature remains below that obtained from the observed HD48099 spectrum. For HD206267, a contribution of 20 per cent gives a surface gravity estimate close to the IACOB value for HD199579, whereas larger contributions approach the primary-component surface gravity published by \citet{Raucq2018}. None of these simulated spectra reproduces the estimated effective temperature of the observed HD206267 spectrum. These comparisons alone are insufficient to select a unique contribution of the secondary component. Estimates and uncertainties from separate calculations have not been combined into a simultaneous fit of the two physical parameters.

The diagnostic-line comparisons in \ref{app:line_profiles} complement the parameter comparisons for HD48099 and HD206267. The full optical comparison in \ref{app:full_spectra} is shown for HD48099 only. Agreement in a few prominent lines does not imply agreement in all 34 equivalent widths used by the network. Similarly, close estimates of effective temperature and surface gravity do not ensure matching profiles throughout the optical spectrum. These spectral comparisons are therefore needed to assess differences that the two estimated parameters alone do not describe.

The main simulations add observed spectra without imposing an additional relative Doppler shift. The profiles in \ref{app:velocity} illustrate how shifting the secondary changes the shape of the combined absorption lines. The scaling factor is a controlled numerical input. It does not represent a physical luminosity ratio derived from stellar radii, distances, or evolutionary constraints. Orbital phase, rotation, variability, stellar winds, and wavelength-dependent relative fluxes can also affect the comparison with an observed binary spectrum. The available training database is restricted to single O-type stars and the optical region, which further limits the interpretation of these results.

Future work should include a larger sample of observed binary systems and simulated spectra with these additional physical effects. A network trained with binary spectra could then be evaluated for the estimation of the individual component parameters and their relative contributions. This analysis would require comparisons with systems for which the orbital and atmospheric parameters have been determined independently. These developments would improve the analysis of massive-star populations that include unresolved binaries.

\section{Conclusions}
\label{sec:conclusions}

We used the recurrent neural network of \citet{Flores2023} to estimate effective temperature and surface gravity from spectra representative of unresolved O-type binaries. The network was developed for single O-type stars and uses 34 equivalent widths. We tested two approximations: adding a smooth blackbody continuum and combining two observed IACOB spectra. The principal conclusions are:

\begin{enumerate}
\item The archived blackbody cases give closely spaced estimates of $T_{\rm eff}$ and $\log g$. Adding the continuum can produce larger differences from the estimates for the unmodified spectrum.
\item Adding an observed secondary spectrum can produce much larger differences than those among the archived blackbody cases. The companion alters the diagnostic lines as well as the continuum.
\item For the general comparison of 192 simulated systems and 1,152 modified spectra, the effective temperature and surface gravity estimates are predominantly below those obtained for the isolated primary spectra. The individual examples show that this tendency does not hold for every combination or percentage contribution.
\item At 75 per cent secondary contribution, the median differences in estimated effective temperature and surface gravity are smaller in magnitude for pairs at or below each of the four tested catalog effective-temperature cutoffs. The distributions overlap, and the cutoff is not a physical boundary.
\item The individual comparisons depend on both the physical parameter and the published reference. For HD190429, different contributions give the closest effective temperature and surface gravity to the IACOB values. The 5 per cent HD48099 simulation approaches the values attributed to \citet{Soloveva_1986}. For HD206267, none of the six simulations reproduces the estimated effective temperature of the observed spectrum. The proposed method is sensitive to a secondary contribution, but these results do not establish a unique contribution or separate an observed binary spectrum into its components.
\end{enumerate}

This analysis provides a starting point for future work with a larger sample of observed binary systems and simulated spectra that include different relative radial velocities, orbital phases, and component fluxes. Future models could identify composite spectra and estimate the properties of both components simultaneously, improving automated spectroscopic analysis of large surveys.

\balance
\bibliography{references}

% The diagnostic spectra need full-width panels. Start the appendices at full
% width so their explanatory paragraphs and captions stay with the spectra.
\onecolumn
\appendix
%\makeatletter
%\@addtoreset{figure}{section}
%\makeatother
\renewcommand{\theHfigure}{\Alph{section}.\arabic{figure}}

\section{Representative spectra for the blackbody-continuum experiment}
\label{app:blackbody_spectra}

\begin{figure}[H]
\centering
\subfloat[HD14947.]{\includegraphics[width=0.82\linewidth]{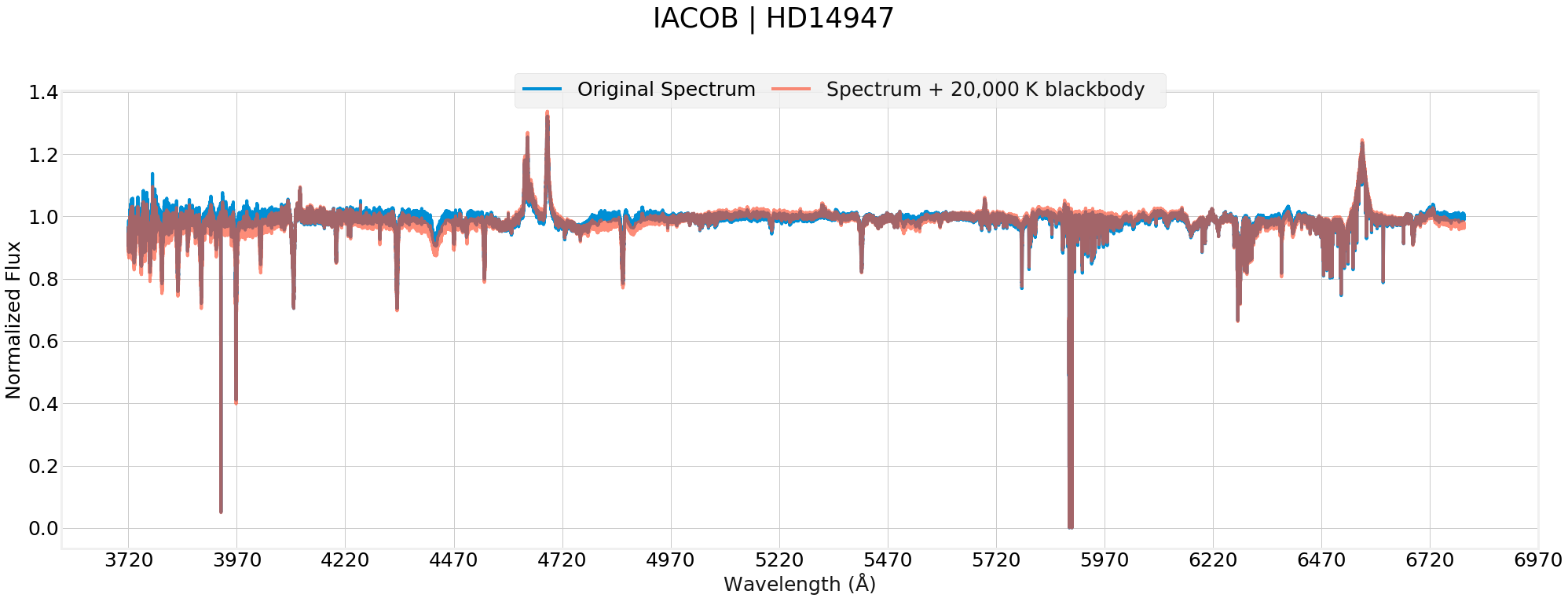}}
\\
\subfloat[HD15570.]{\includegraphics[width=0.82\linewidth]{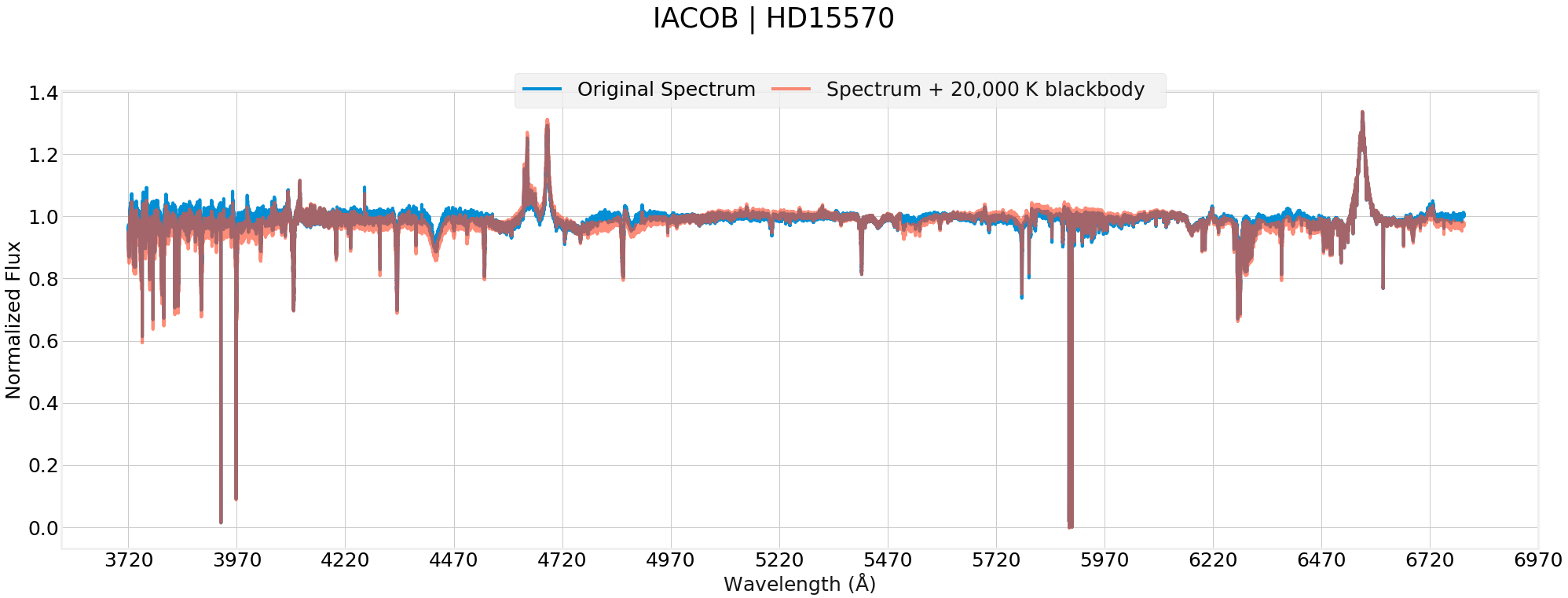}}
\\
\subfloat[HD34078.]{\includegraphics[width=0.82\linewidth]{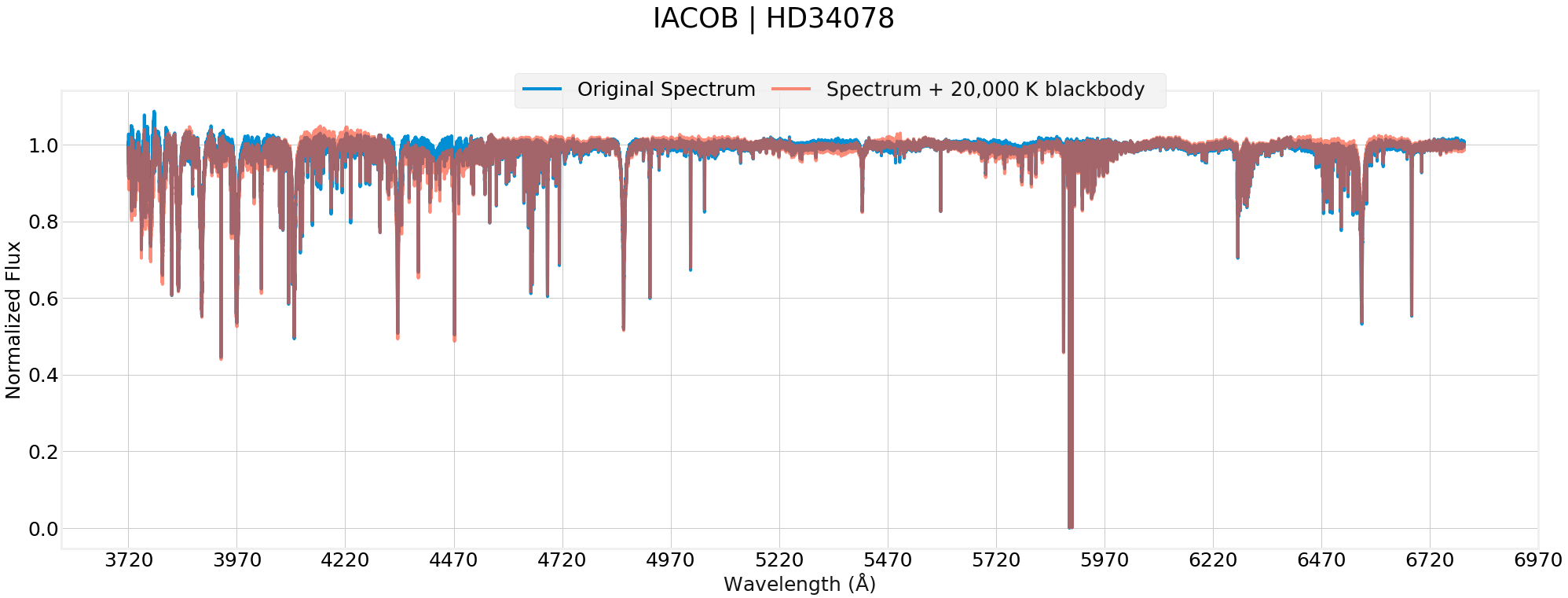}}
\caption{Observed spectra before and after adding a 20,000~K blackbody continuum and renormalizing the combined signal.}
\label{fig:blackbody_spectra}
\end{figure}

Figure~\ref{fig:blackbody_spectra} compares the observed spectra of HD14947, HD15570, and HD34078 with the spectra obtained after adding a 20,000~K blackbody continuum and renormalizing the combined signal. These examples provide the detailed spectral comparison for the procedure in Section~\ref{sec:blackbody_method}; the corresponding estimates of effective temperature and surface gravity are discussed in Section~\ref{sec:results} and shown in Figure~\ref{fig:blackbody_results}.

\clearpage
\section{Representative line-profile comparisons}
\label{app:line_profiles}
\begin{figure}[H]
\centering
\subfloat[Blue wavelength group.]{\includegraphics[width=0.85\linewidth]{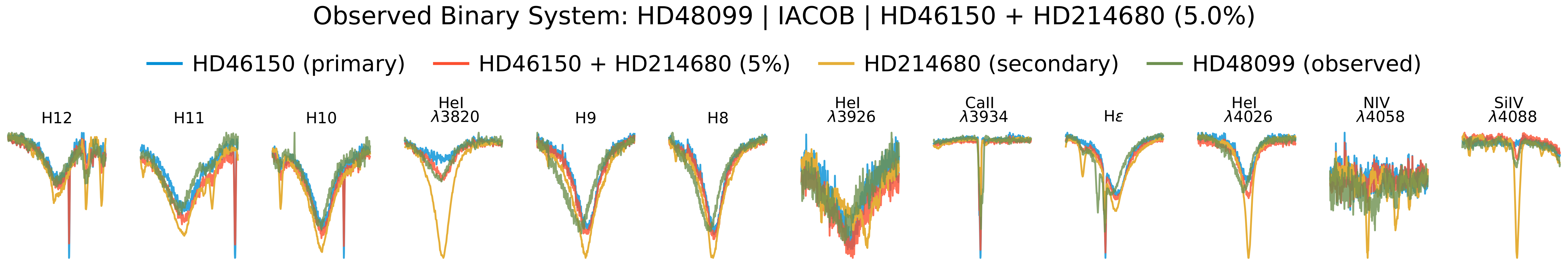}}
\\
\subfloat[Central wavelength group.]{\includegraphics[width=0.85\linewidth,trim=0 0 0 38,clip]{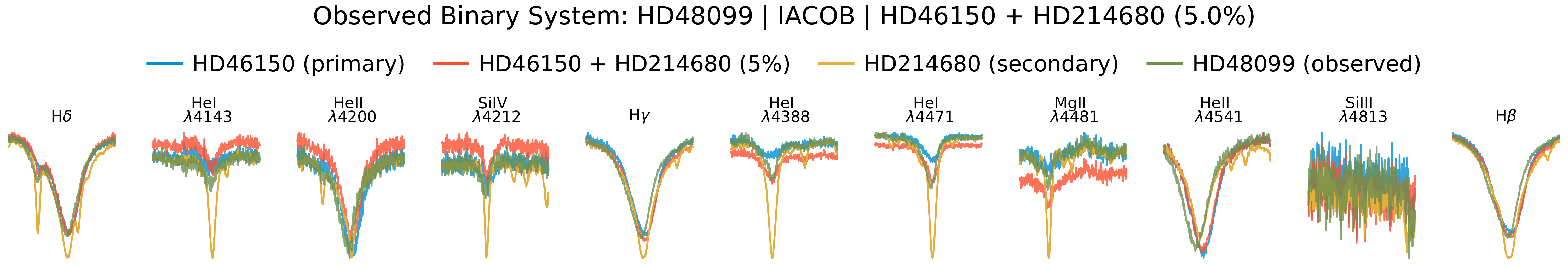}}
\\
\subfloat[Red wavelength group.]{\includegraphics[width=0.85\linewidth,trim=0 0 0 38,clip]{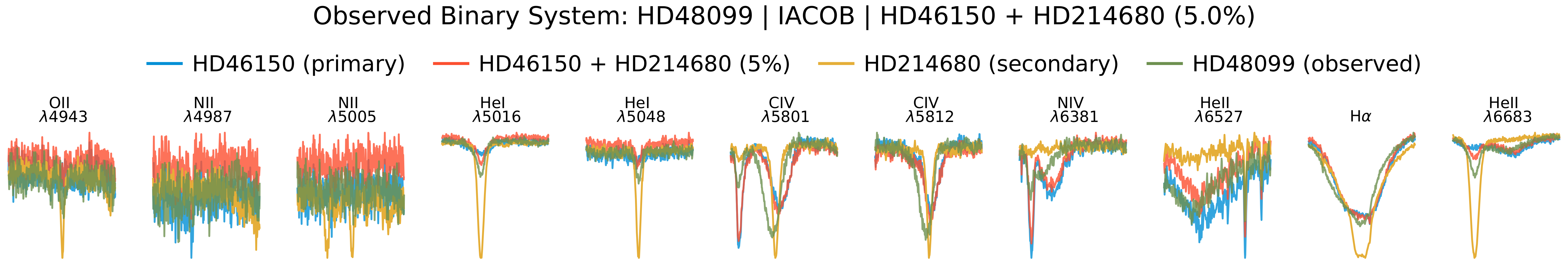}}
\caption{Diagnostic lines for observed HD48099 and the HD46150+HD214680 simulation with a 5 per cent secondary contribution. The legend applies to all three groups.}
\label{fig:hd48099_lines}
\end{figure}

\begin{figure}[H]
\centering
\subfloat[Blue wavelength group.]{\includegraphics[width=0.85\linewidth]{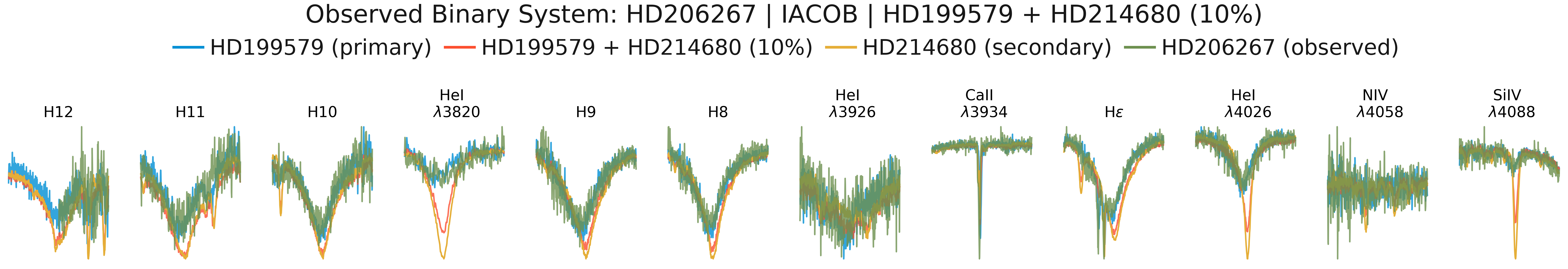}}
\\
\subfloat[Central wavelength group.]{\includegraphics[width=0.85\linewidth,trim=0 0 0 150,clip]{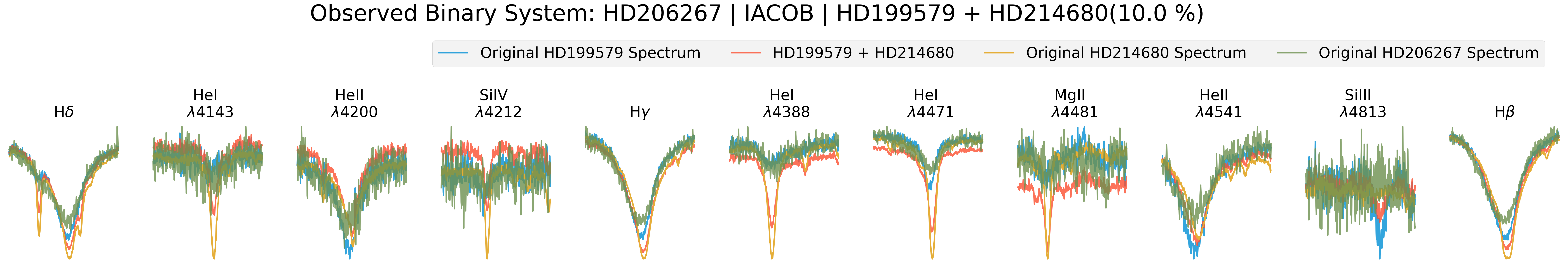}}
\\
\subfloat[Red wavelength group.]{\includegraphics[width=0.85\linewidth,trim=0 0 0 150,clip]{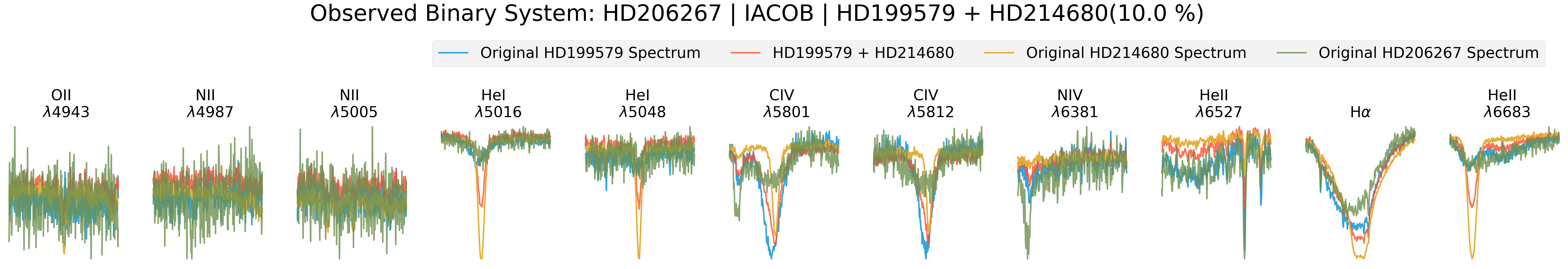}}
\caption{Diagnostic lines for observed HD206267 and the HD199579+HD214680 simulation with a 10 per cent secondary contribution. The legend applies to all three groups.}
\label{fig:hd206267_lines}
\end{figure}

Figures~\ref{fig:hd48099_lines} and \ref{fig:hd206267_lines} compare diagnostic lines of the observed binary systems, the simulated binary spectra, and the isolated spectra of the stars used in each simulation. The three wavelength groups show examples with a contribution of 5 per cent for HD48099 and 10 per cent for HD206267. These examples illustrate the line profiles at the specified contributions; they do not establish a simultaneous fit of effective temperature and surface gravity. For HD48099, the 5 per cent example allows the line profiles to be examined alongside the parameter comparison with \citet{Soloveva_1986}. The HD206267 example at 10 per cent is a separate illustration of the addition procedure; it is not a contribution selected by agreement with both estimated parameters.

\clearpage
\section{Full-spectrum comparison}
\label{app:full_spectra}

\begin{figure}[H]
\centering
\subfloat[3700--4500~\AA.]{\includegraphics[width=0.86\linewidth]{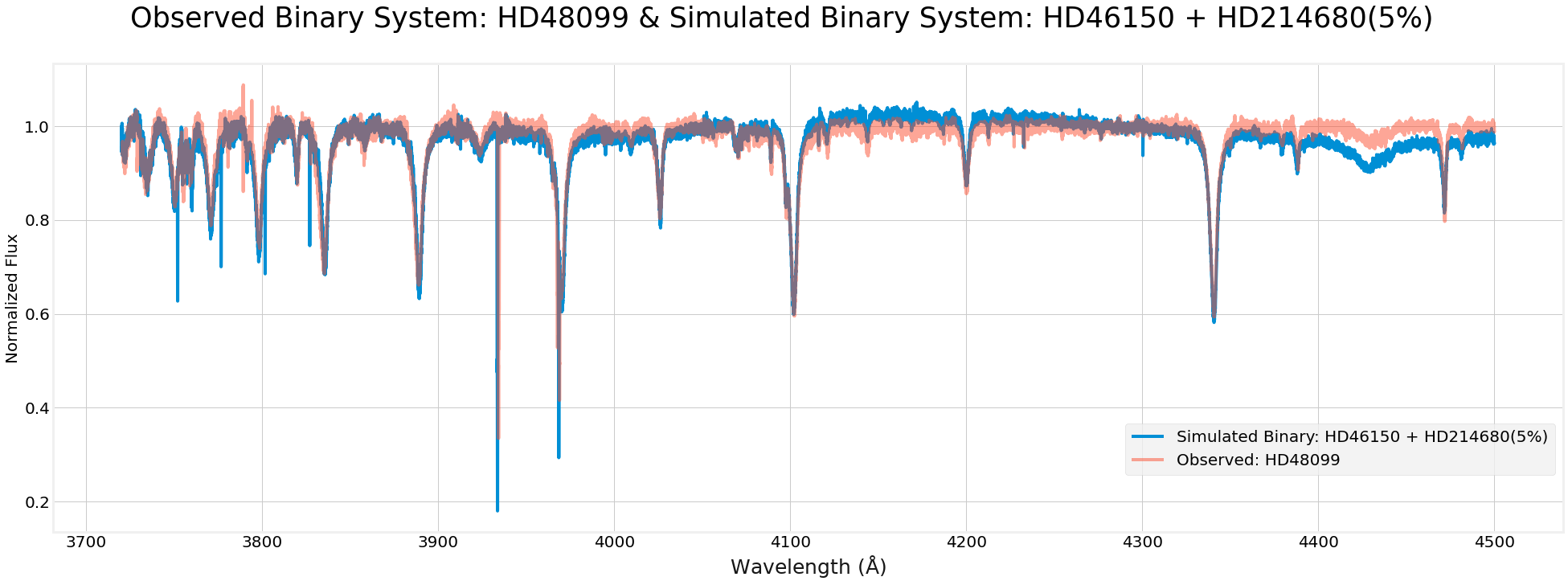}}
\\
\subfloat[4500--5500~\AA.]{\includegraphics[width=0.86\linewidth]{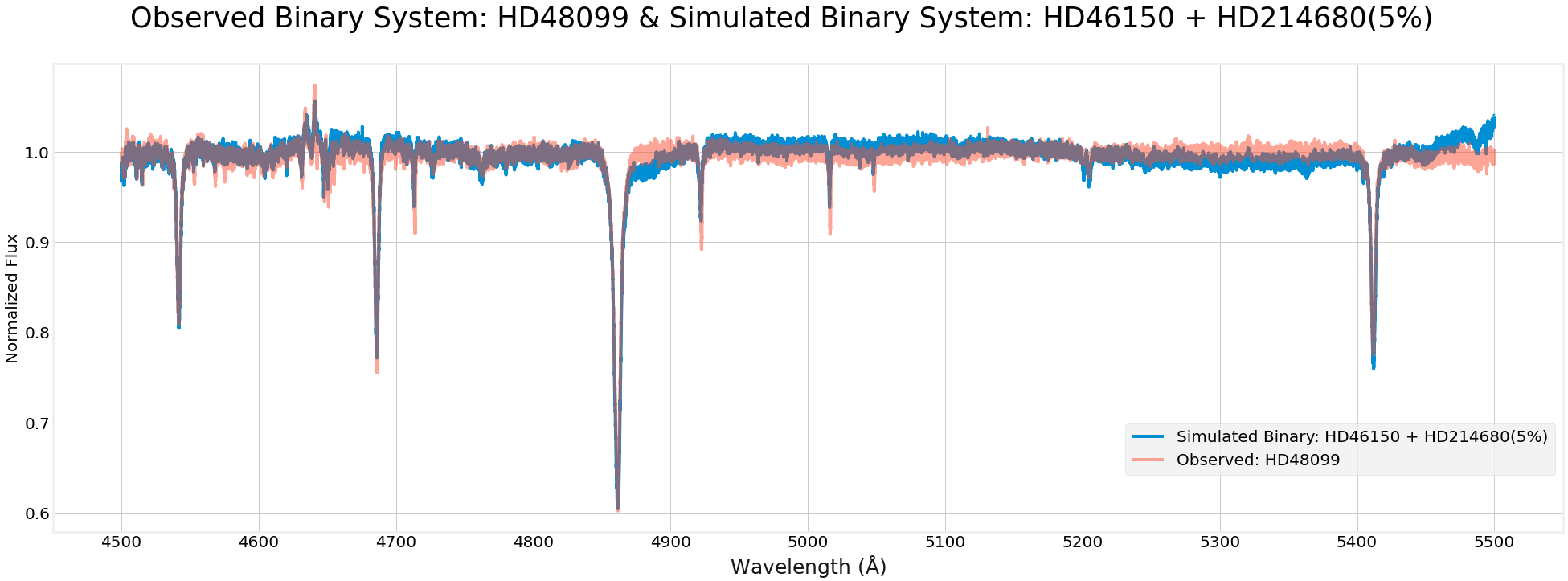}}
\\
\subfloat[5500--6800~\AA.]{\includegraphics[width=0.86\linewidth]{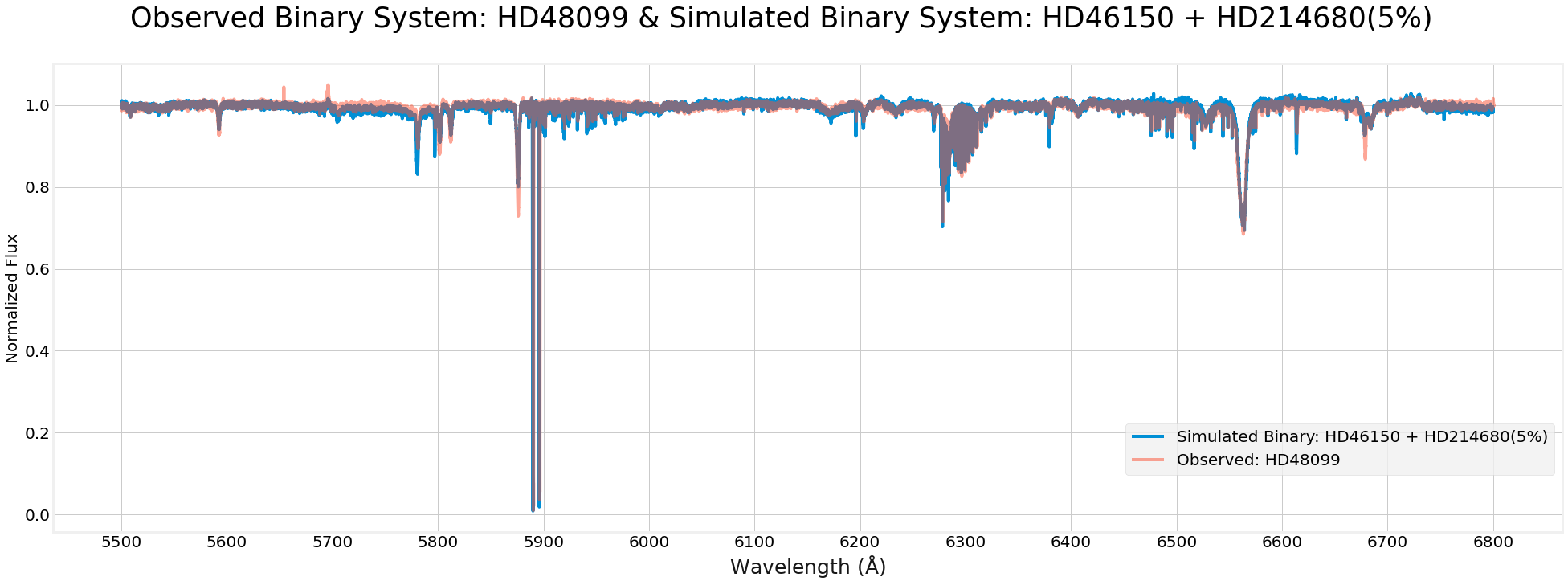}}
\caption{Observed HD48099 compared with the HD46150+HD214680 simulation at 5 per cent secondary contribution.}
\label{fig:hd48099_full}
\end{figure}

The full optical comparison for HD48099 in Figure~\ref{fig:hd48099_full} shows that the simulated and observed spectra can agree closely in several strong lines while retaining differences elsewhere. Agreement in a few prominent lines does not require agreement in all 34 equivalent widths used by the network. Conversely, similar estimated effective temperature and surface gravity do not guarantee the same profiles throughout the spectrum. The full-spectrum comparison is therefore complementary to the parameter comparison and cannot be replaced by agreement in the two estimated values alone.

\clearpage
\section{Adding Doppler-shifted absorption spectra}
\label{app:velocity}

\begin{figure}[H]
\centering
\includegraphics[width=\textwidth]{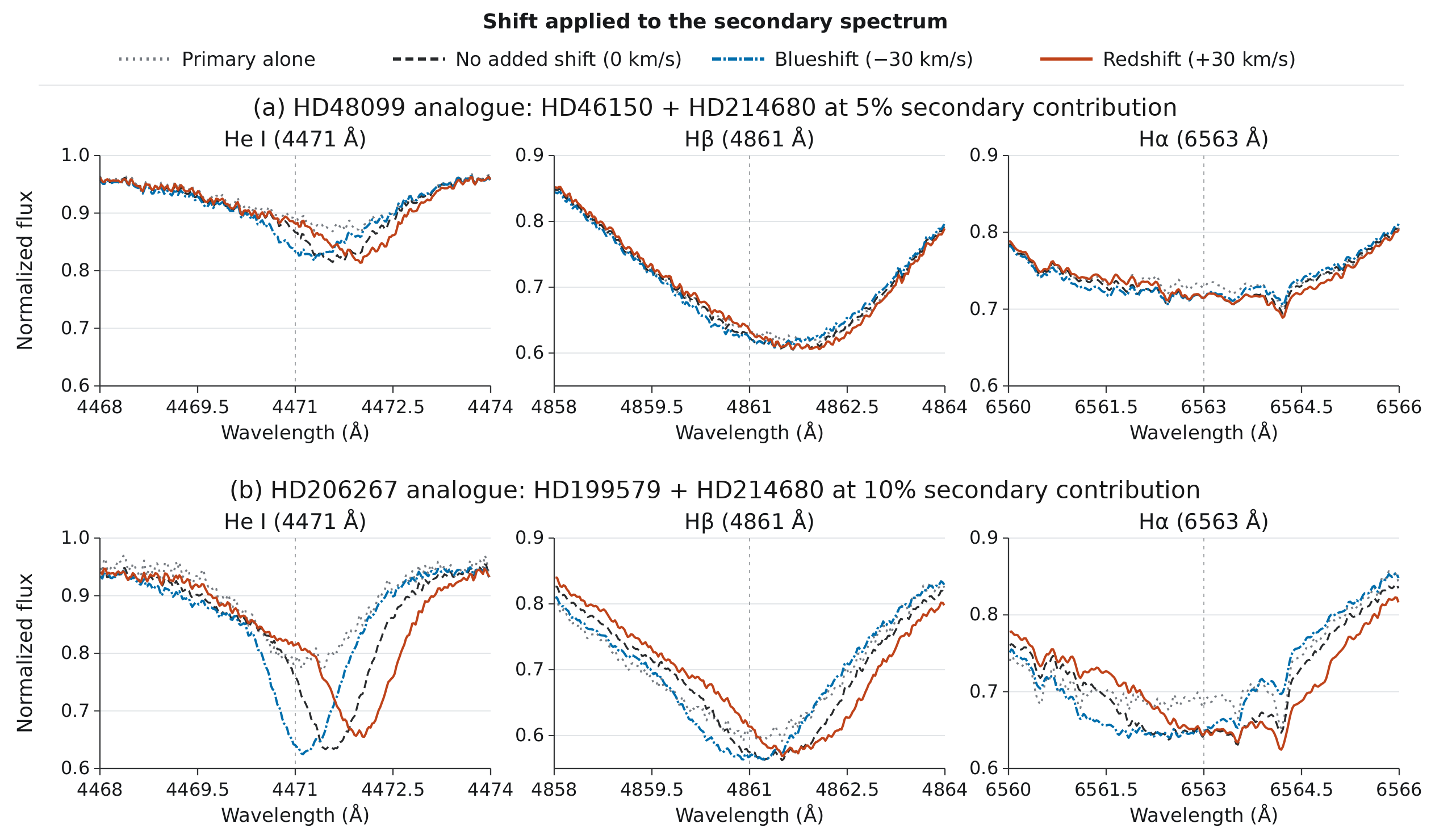}
\caption{Normalized flux versus wavelength around He\,I $\lambda4471$, H$\beta$, and H$\alpha$. The primary-only spectrum is shown alongside composites formed by adding HD214680 with an imposed blueshift ($-30~\mathrm{km\,s^{-1}}$), no added shift, or an imposed redshift ($+30~\mathrm{km\,s^{-1}}$). Each panel shows a 6~\AA\ window around the labeled wavelength, marked by a vertical dotted line.}
\label{fig:velocity_profiles}
\end{figure}

Figure~\ref{fig:velocity_profiles} shows three absorption lines for the analogue pairs and scaling factors in \ref{app:line_profiles}. We sample the secondary flux at $\lambda/(1+\Delta v/c)$ for each primary-grid wavelength $\lambda$ before adding it as in Equation~\ref{eq:binary_sum}. All shifts use common wavelength coverage. We normalize the composites and the primary-only spectrum by the same procedure.

The $0.66$~\AA\ H$\alpha$ displacement applies to the secondary line. The primary stays fixed, and the shifted secondary changes the shape of the blend. The FITS spectra were not barycentric-corrected, so the velocities label shifts imposed on the recorded spectra.

\end{document}